\documentclass[aps,pra,twocolumn,superscriptaddress,floatfix,nofootinbib,showpacs,longbibliography
]{revtex4-2}
\usepackage{soul}
\usepackage{xcolor}
\usepackage{graphicx}% Include figure files
\usepackage{amsthm, amssymb, mathtools}
\usepackage{dcolumn}% Align table columns on decimal point
\usepackage{bm}% bold math
\usepackage{braket}
\newtheorem{theorem}{Theorem}
\newtheorem{lemma}{Lemma}
\newtheorem{corollary}{Corollary}
\usepackage[utf8]{inputenc}
\usepackage[T1]{fontenc}
\usepackage{mathptmx}

\usepackage{mathtools,xcolor}
\usepackage{hyperref}

\usepackage{stmaryrd}
\usepackage{soul}
\usepackage{verbatim}
\usepackage{amsmath}
\usepackage{amssymb}
\usepackage{epsfig}
\usepackage{color}
\usepackage{amsthm}
\usepackage{dsfont}

\newcommand{\Tr}{\mathrm{Tr}}
\newcommand{\cE}{\mathcal{E}}
\newcommand{\cF}{\mathcal{F}}
\newcommand{\cN}{\mathcal{N}}
\newcommand{\bR}{\mathbb{R}}

\newcommand{\cH}{\mathcal{H}}

\newtheorem{definition}{Definition}
\newtheorem{proposition}{Proposition}
\newtheorem{fact}{Fact}

\begin{document}
	
\title{A finite sufficient set of conditions for catalytic majorization}	

\author{David Elkouss}
 \email{david.elkouss@oist.jp}
\author{Ananda G. Maity}
\email{anandamaity289@gmail.com}
\affiliation{Networked Quantum Devices Unit, Okinawa Institute of Science and Technology Graduate University, Onna-son, Okinawa 904-0495, Japan}
\author{Aditya Nema}
\email{aditya.nema30@gmail.com}
\affiliation{Institute of Quantum Information, RWTH Aachen, Germany}
\author{Sergii Strelchuk}%
 \email{sergii.strelchuk@cs.ox.ac.uk\\ All authors have contributed equally in this work.}
\affiliation{
Department of Computer Science,
University of Oxford
}%

             
%%%%%%%%%%%%%%%%%%%%%%%%%%%%%%%%%%%%%%%%%%%%%%
\begin{abstract}
The majorization relation has found numerous applications in mathematics, quantum information and resource theory, and quantum thermodynamics, where it describes the allowable transitions between two physical states. In many cases, when state vector $\mathbf x$ does not majorize state vector $\mathbf y$, it is nevertheless possible to find a catalyst -- another vector $\mathbf z$ such that $\mathbf x\otimes \mathbf z$ majorizes $\mathbf y\otimes \mathbf z$. Determining the feasibility of such catalytic transformation typically involves checking an infinite set of inequalities. Here, we derive a finite sufficient set of inequalities that imply catalysis. Extending this framework to thermodynamics, we also establish a finite set of sufficient conditions for catalytic state transformations under thermal operations. For novel examples, we provide a software toolbox implementing these conditions. 

\end{abstract}
%%%%%%%%%%%%%%%%%%%%%%%%%%%%%%%%%%%%%%%%%%%%%%
\maketitle
\section{Introduction}\label{sec:intro}
Irreversibility is a phenomenon in the realm of physics that serves as a key indicator for transitions from orderly to disorderly states of the systems. It is typically described by the laws of thermodynamics, yet its relevance extends far beyond thermodynamics, influencing diverse fields across physics and mathematics. Intuitively, a process is considered reversible when no dissipation occurs, allowing the system to remain in equilibrium throughout its evolution. Accurately characterizing irreversibility is thus paramount, with important applications in areas such as quantum state transformations, entanglement dynamics, characterization of thermal machines and heat engines.

There is a broader notion of irreversibility which extends beyond thermodynamic irreversibility, treating it as a special case. Central to this approach is the concept of {\it majorization}, which provides a powerful tool for distinguishing between reversible and irreversible processes.

Majorization and its implications had a major impact on various scientific disciplines such as mathematics~\cite{marshall1979inequalities}, machine learning~\cite{mairal2015incremental, sun2016majorization}, and economics~\cite{marshall1979inequalities}. In recent times, it has played a crucial role in quantum information theory~\cite{nielsen1999conditions,jonathan1999entanglement,pereira2015extending,kribs2013trumping,pereira2013dirichlet}, quantum  thermodynamics~\cite{brandao2015second,Horodecki2013,Gour2018}, and more generally on resource theories \cite{fritz2017resource}. See \cite{marshall1979inequalities} for a review on early applications. In the latter cases, it gives insights into the entanglement structure of quantum states, thermodynamic resources and provides an elegant operational criterion which determines when it is possible to convert one quantum state into another.

Nevertheless, majorization alone falls short of offering a comprehensive characterization of state transformations. There are instances where transforming one quantum state into another is not possible within a specific class of operations. Yet, the introduction of an ancillary state can sometimes facilitate this transformation under the same operations, with the ancillary state ultimately returning to its original and exact form \cite{jonathan1999entanglement}. This additional system that alleviates restrictions on apparently prohibited transitions serves as a catalyst. For a comprehensive review of different aspects of catalysis in quantum information theory, we refer interested readers to the following references \cite{Datta_2023,lipkabartosik_2023}.

To illustrate further, let us analyze this from the perspective of thermodynamics. In the macroscopic domain of thermodynamics, a system existing in state $\rho$ can undergo a transition to state $\sigma$ if there is a decrease in free energy where the free energy of a state $\rho$ is defined as, $\cF(\rho) = \langle E(\rho)\rangle - KTS(\rho)$. Here $\langle E(\rho)\rangle$ is the average energy, $K$ is the Boltzmann constant, and $S(\rho)$ is the entropy of the state $\rho$ with $T$ being the temperature of the surrounding heat bath. Unfortunately, in the realm of microscopic, quantum, or highly correlated systems, a reduction in free energy alone does not suffice as a condition for state transformation. Instead, the concept known as {\it thermo-majorization} has been introduced \cite{brandao2015second,Lostaglio_review_19}. Nevertheless, even thermo-majorization proves insufficient for characterizing state transformation rules comprehensively. Instances exist where two states $\rho$ and $\sigma$ do not adhere to thermo-majorization, yet the transition from $\rho$ to $\sigma$ becomes feasible with the presence of an auxiliary system $\chi$, known as catalyst which remains unaltered before and after the state transitions \cite{Lostaglio_review_19}.

Therefore, when assessing the possibility of transforming $\rho$ into $\sigma$, consideration must be given to the existence of another working body or other ancillary systems $\chi$ such that the composite system $\rho \otimes \chi$ can be transformed into $\sigma \otimes \chi$. Thus, majorization should be exclusively applied to the total resources, encompassing all potential catalysts and working bodies of arbitrary dimensions, rather than to the system of interest itself. Criteria that investigate transitions between two states in the presence of catalysis are commonly termed as {\it trumping}. In \cite{brandao2015second}, it has been demonstrated that merely decreasing free energy is insufficient to fully encapsulate state transformation rules under thermal operations. Instead (for states diagonal in the energy eigenbasis), the necessary and sufficient conditions for thermodynamic transitions are determined by a family of second laws, which require verifying the increase of the generalized free energy, i.e., $\cF_p(\rho, H) \geq \cF_p(\sigma, H) $ for $p\in (-\infty, \infty)$. A formal definition of generalized free energy and a comprehensive discussion on these conditions are presented in Section IV.

Verifying the existence of catalytic majorization or trumping relationships requires evaluating an infinite number of conditions across the continuous spectrum of $p$ values. Despite having well-defined mathematical expressions for these catalytic transformation conditions, checking an unbounded number of inequalities is computationally infeasible. A fundamental question is whether these infinite conditions can be reduced to a finite, tractable set of conditions that guarantee catalytic majorization. 

Prior research has approached this problem from multiple angles. For probability distributions of pure states or $d \times d$ mixed states where $d\geq 4$, Gour \cite{Gour05} established that a finite number of conditions cannot be sufficient to determine transformation feasibility. Existing characterizations of necessary and sufficient conditions, such as those developed by Klimesh \cite{klimesh2007inequalities}, Turgut \cite{turgut2007catalytic} and Aubrun and Nechita \cite{aubrun2008catalytic}, all require infinite checks and lack computational tractability. These limitations have directed research efforts toward approximate catalytic transformations \cite{Condra, Wilming}, with relatively little progress on exact catalytic transformations.

\begin{figure}[t!]
\centering
    \includegraphics[width=8cm]{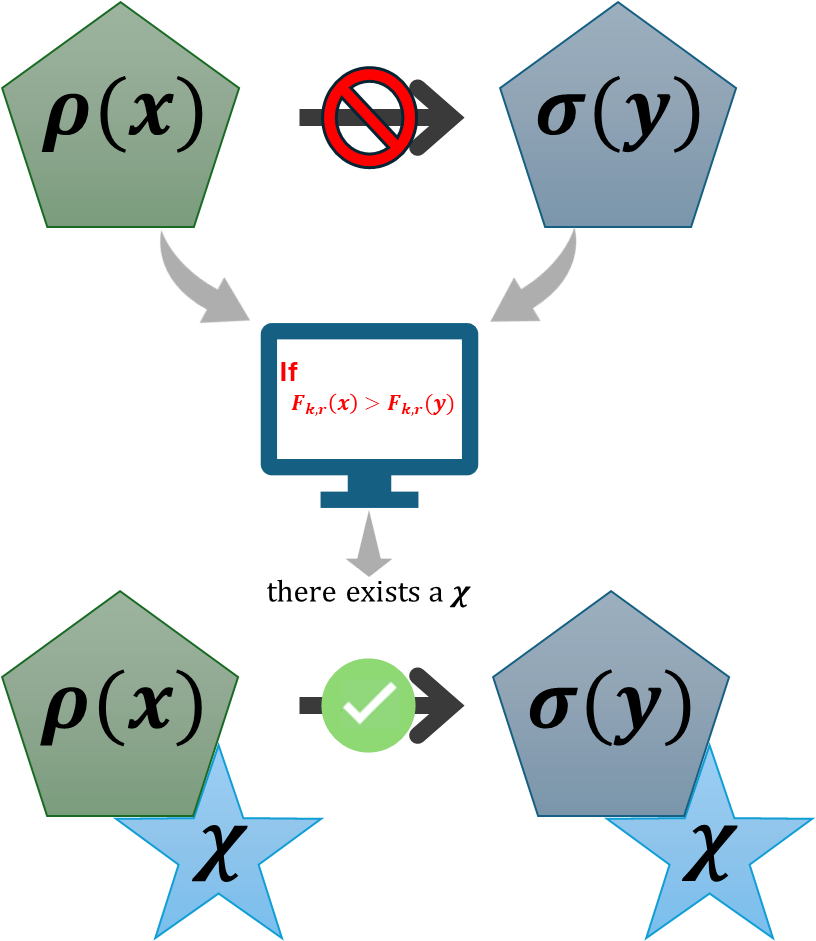}
        \caption{Checking a finite set of sufficient criteria whether there exist a catalysis $\chi$ that can alleviate restrictions on apparently prohibited transitions from state $\rho$ to state $\sigma$ having eigenvalues $\mathbf x$ and $\mathbf y$ respectively.}
    \label{Fig1}
\end{figure}

In this work, we derive a finite set of sufficient conditions to determine whether state transformation is possible under a given free operation (as illustrated in Fig.~\ref{Fig1}). We first derive this finite set of conditions which are sufficient for trumping two state vectors $x_{AB}$ and $y_{AB}$ shared between two spatially separated parties under local operation and classical communication (LOCC). Further, we also extend our analysis to the second law of thermodynamics and again explicitly derive a finite set of sufficient conditions for trumping states that are block diagonal in the energy eigenbasis via thermal operations. Our analysis also applies to a broader class of general states that are not necessarily block-diagonal in the energy eigenbasis. Finally, we introduce a simulation toolbox to facilitate the practical application of our trumping criteria.

%%%%%%%%%%%%%%%%%%%%%%%%%%%%%%%%%%%%%%%%%%%%%%
\section{Majorization and Catalytic majorization}\label{s2}
We start by reviewing the conditions for transformation between two general state vectors shared between two distant parties under LOCC (i.e. within the realm of entanglement theory) and subsequently explore how these conditions are altered in the presence of catalysis.

Consider a vector in the Hilbert space $y_{AB} \in \cH_{AB} (=\cH_A \otimes \cH_B)$ that represents some target state that two spatially separated parties, Alice and Bob, wish to share by acting on the available state $x_{AB}\in \cH_{AB}$ using LOCC~\cite{bennett1996concentrating}. More formally, recall that any such bipartite quantum state in the tensor product space $ \cH_{AB}$ can be written in the Schmidt form~\cite{peres2006quantum}:
\begin{equation}
  x_{AB}=\sum_{i=1}^{n} \sqrt{ x_i}{u_i}_A\otimes{v_i}_B
\end{equation}
where
$\{{u_i}_A\}_i \in \cH_A$ and $\{{v_i}_B\}_i \in \cH_B$
are two sets of orthonormal bases and $|\cH_{AB}|=|\cH_A||\cH_B| = n$. The vector $\mathbf x=(x_1,\ldots,x_n)\in P_n$, with $P_n=\{(x_1,\ldots,x_n): x_1\geq x_2\geq\ldots x_n\geq 0,\; \sum_i x=1\}$, is called the vector of Schmidt coefficients. Abusing notation, we will sometimes denote the state by its vector of Schmidt coefficients. 

Here we restrict our attention to deterministic transformations. In this case, it was shown by Nielsen \cite{nielsen1999conditions} that given $\mathbf x,\mathbf y \in P_n$,  $\textbf x$ can be transformed into $\textbf y$ if and only if for all $i\in\{1,\ldots,n\}$
\begin{equation}
    \sum_{j=1}^i x_j\leq\sum_{j=1}^i y_j.
\end{equation}
This relation is known as majorization. Here we say that $\mathbf y$ majorizes $\mathbf x$ or equivalently that $\mathbf x$ is majorized by $\mathbf y$ and we compactly denote it by $\mathbf x \prec \mathbf y$. Majorization is a partial order, i.e. there exist vectors $\mathbf x, \mathbf y$ for which neither $\mathbf x\prec\mathbf y$ nor $\mathbf y\prec \mathbf x$. We indicate that $\mathbf x$ is not majorized by $\mathbf y$ by $\mathbf x\nprec \mathbf y$.

In some cases, when $\mathbf x \nprec \mathbf y$, it is nevertheless possible to convert $x_{AB}$ to $y_{AB}$: surprisingly, there may exist a special vector that acts as a catalyst, i.e. there exists a set of triples of states with ($ \mathbf x,\mathbf y,\mathbf z$) such that $\mathbf y$ does not majorize $\mathbf x$, but $\mathbf y\otimes \mathbf z$ majorizes $\mathbf x\otimes \mathbf z$. In other words, by using an auxiliary state $z_{AB}$ with vector $\mathbf z$ one enables the deterministic transformation, and at the end of the latter $z_{AB}$  is recovered unscathed. This relation is called catalytic majorization or trumping \cite{jonathan1999entanglement} and henceforth we shall denote it by $\mathbf x\prec_T \mathbf y$.

We define by $T(\mathbf y)=\{\mathbf x\in P_{<\infty}:  \mathbf x\prec_T \mathbf y\}$ the set of vectors that are trumped by $\mathbf y$, where $P_{<\infty}=\cup_{n=1}^\infty P_n$. Note that if $\text{dim}(\mathbf x)\neq\text{dim}(\mathbf y)$, the definition of trumping and majorization can be extended to probability vectors where the shorter vector is padded with zeros.

The first comprehensive set of conditions characterizing the trumping relation was derived in 2007 independently by Klimesh and Turgut~\cite{turgut2007catalytic,klimesh2007inequalities} (see also  \cite{muller2016generalization} for a characterization of a generalized concept of majorization). Remarkably, although the catalyst may be of arbitrary dimension, these conditions are equivalent to all the following strict inequalities being true \cite{muller2016generalization}: 
\begin{equation}
\label{eq:condt}
 \mathbf x\in T(\mathbf y) \iff
   \begin{cases}
   H_p (\mathbf x) >H_p (\mathbf y),& \text{for}~p\in\mathbb R/\{0\}\\
   H_{\text{burg}}(\mathbf x) > H_{\text{burg}}(\mathbf y) 
   \end{cases}
\end{equation}
where for $p\neq\{0,1\}$:
\begin{equation}
    H_p (\mathbf x) :=\frac{\text{sign}(p)}{1-p}\log\sum_{i=1}^n(x_i)^{p}\ ,
\end{equation}
for $p=1$:
\begin{equation}
H_1(\mathbf x) := -\sum_{i=1}^nx_i\log x_i ,
\end{equation}
and
\begin{equation}
H_{\text{burg}}(\mathbf x):=\frac{1}{n}\sum_{i=1}^n \log x_i\ .
\end{equation}
Note that both $H_p$ for $p<0$ and $H_{\text{burg}}$ evaluate to $-\infty$ if some of the elements of $\mathbf x$ are zero.  

The closure of $\overline{T(\mathbf y)}$ in the $l_1$-norm was introduced in \cite{aubrun2008catalytic} as a proxy for the study of $T(\mathbf y)$ (see also \cite{aubrun2009stochastic}). It turns out that by including the limit points of $T(\mathbf y)$, one can simplify the set of conditions for trumping from Eq.~\eqref{eq:condt} at the expense of adding `non-physical' catalysts:
\begin{equation}
\label{eq:condtbar}
 \mathbf x\in \overline{T(\mathbf y)} \iff
   H_p (\mathbf x) >H_p (\mathbf y),  p\in(1,\infty).
\end{equation}
$\overline{T(\mathbf y)}$ can be interpreted as the set of vectors that are approximately trumped by $\mathbf y$.
In the following, we rephrase the characterization of $\overline{T(\mathbf y)}$ and ${T(\mathbf y)}$ in a more convenient form for the purpose of this paper. Let $\mathbf x\in\mathbb R_+^n$ and $p\in\mathbb R$, with slight abuse of notation, we define
\begin{equation} \label{eq:scalednorm_p}
    \|\mathbf x\|_p:=\left(\frac{1}{n}\sum_{i=1}^n(x_i)^p\right)^{1/p}\ .
\end{equation}
Note that $\|\cdot\|_p$ is only a norm (up to a factor $n^{-1/p}$, the Schatten $p$-norm) for $p\geq 1$. We call the weight of a vector the number of non-zero entries. We set $\|\mathbf x\|_0=\left(\prod_{i=1}^nx_i\right)^{1/n}$ and for $p<0$ and $\mathbf x$ with weight smaller than $n$, $\|\mathbf x\|_p=0$ such that $\|\cdot\|_p$ is continuous. With this notation we can replace Eq.~\eqref{eq:condt} and Eq.~\eqref{eq:condtbar} with:

\begin{equation}
\label{eq:condtnorm}
 \mathbf x\in T(\mathbf y) \iff
   \begin{cases}
   \|\mathbf x\|_p < \|\mathbf y\|_p, & p\in(1,\infty) \\
   \|\mathbf x\|_p > \|\mathbf y\|_p, & p\in(-\infty,1) \\
   H_1(\mathbf x) > H_1(\mathbf y),
   \end{cases}
\end{equation}

and

\begin{equation}
\label{eq:condtbarnorm}
 \mathbf x\in \overline{T(\mathbf y)} \iff
   \begin{cases}
   \|\mathbf x\|_p < \|\mathbf y\|_p, & p\in(1,\infty).
   \end{cases}
\end{equation}

%%%%%%%%%%%%%%%%%%%%%%%%%%%%%%%%%%%%%%%%%%%%%%

\section{A finite sufficient set of conditions for trumping of state vectors} \label{s3}

Despite Eq.~\eqref{eq:condtnorm} (equivalently \eqref{eq:condt}) and Eq.~\eqref{eq:condtbarnorm} (equivalently \eqref{eq:condtbar}) respectively form a complete set of the necessary and sufficient conditions for trumping and approximate trumping, the number of conditions to check remains infinite. Here, we aim to outline our primary mathematical findings that lead to a finite set of conditions, which guarantee the trumping of bipartite state vectors when the free operation is LOCC.

For the concrete case of four-dimensional vectors and two-dimensional catalysts, necessary and sufficient conditions are known \cite{sun2005existence,bosyk2018lattice}. 
The general problem of characterizing the existence of a $k$ dimensional catalyst for $n$ dimensional vectors has been answered with algorithms that have running time polynomial in $n$ and $k$~\cite{bandyopadhyay2002efficient,sun2005existence}. However, the algorithm only provides an answer for a given value of $k$.
Some general conditions necessary for trumping were investigated in \cite{sanders2009necessary}. 
Here, we are interested in a different approach to the problem of catalytic majorization. Instead of bounding the dimension of the catalyst, we want to find a discrete set of inequalities that implies the sets in Eqs.~\eqref{eq:condt} and \eqref{eq:condtbar}. 

Partial progress in this direction was made by Klemes \cite{klemes2010symmetric} by considering a family of symmetric polynomials $F_{k,r}:\mathbb R^n\mapsto \mathbb R$ indexed by $k,r\in\mathbb N$:
\begin{equation}
	F_{k,r}(\mathbf x) = \sum_{\substack{(k_1,\ldots,k_n)\text{ s.t. }\\\sum_ik_i=k, \max k_i\leq r}}\prod_{i=1}^n \frac{\left(x_i\right)^{k_i}}{k_i!} .
\end{equation}

Intuitively, this can be thought of as a polynomial generated from a truncated exponential function. We define it formally in Definition~\ref{def:polynomial} in Appendix A. Using this polynomial \cite{klemes2010symmetric} showed that there is a {\it finite} set of inequalities that in turn imply the inequality of $p$ norms for a continuous range of values of $p$. We state it as the following fact:
\begin{fact}[Klemes]
	\label{fact:klemes}
Let $\mathbf x,\mathbf y\in\mathbb R_+^n$ and fix $r\in\mathbb N$. If
\begin{equation}
    F_{k,r}(\mathbf x)\geq F_{k,r}(\mathbf y), \qquad k \in \{r, r+1, \ldots, nr\} \nonumber
\end{equation}
Then
\begin{equation}
    \|\mathbf x\|_p\geq \|\mathbf y\|_p, \qquad p\in(0,1) \nonumber
\end{equation}
and further if $\sum_{i=1}^n x_i =\sum_{i=1}^n y_i$ then
\begin{equation}
        \|\mathbf x\|_p\leq \|\mathbf y\|_p \qquad p\in(1,r+1). \nonumber
\end{equation}
\end{fact}
Note, while the condition $\sum_{i=1}^n x_i =\sum_{i=1}^n y_i$  may seem restrictive, it is trivially satisfied when considering physical systems where $x_i, y_i$ indicate the probability of finding the requisite system in state $i$. 

We now introduce some notation before stating our main result.  
Given $\mathbf x\in\mathbb R^n$, we denote by $x_{\min}$ the minimum non-zero element of $\textbf x$: $x_{\min}=\min\{x_i, 1\leq i\leq n: x_i>0\}$.
We denote by $\mathbf x^p$ the point-wise exponentiation of the vector $\mathbf x$. 
That is: $\mathbf x^p=(x_1^p,\ldots,x_n^p)$ where we follow the convention that $0^p=0$ for $p\in\mathbb R$. Abusing notation, we denote $\mathbf x^{-1}=\frac{1}{\mathbf x}$. 
Finally, let $x\in\mathbb R$, we denote the greatest integer smaller than or equal to $x+1$ by $\bar x=\lfloor x+1\rfloor$.

We make use of Fact~\ref{fact:klemes} to construct a $\it finite$ set of sufficient conditions for checking whether an element belongs to the set $T( \mathbf y)$ (and also for $\overline{T(\mathbf  y)}$) in Theorem \ref{th:main}. However, before proceeding we state the following corollaries and lemma derived from Fact~\ref{fact:klemes} that will be pivotal in the proof of our main Theorem~\ref{th:main}. Fact~ \ref{cor:strictineq} is a slight modification of the Fact~\ref{fact:klemes} replacing inequalities with strict inequalities, and some basic properties of the function $\|\mathbf \cdot\|_p$.
\begin{corollary}
\label{cor:strictineq}
Let $n>1$, $\mathbf x,\mathbf y\in P_n$ and fix $r\in\mathbb N$. If
\begin{equation}
    F_{k,r}(\mathbf x)> F_{k,r}(\mathbf y), \qquad k \in \{r, r+1, \ldots, nr\} \nonumber
\end{equation}
Then
\begin{equation}
    \|\mathbf x\|_p> \|\mathbf y\|_p, \qquad p\in(0,1) \nonumber
\end{equation}
and 
\begin{equation}
        \|\mathbf x\|_p< \|\mathbf y\|_p, \qquad p\in(1,r+1). \nonumber
\end{equation}
\end{corollary}

The point-wise notation introduced in the previous section allows us to make a series of identities between different values of $p$. In particular, let $\mathbf x,\mathbf y\in P_n$. We have that for $p,m\in\mathbb R_+$ and $p,m\neq 0$:
\begin{align}
	\|\mathbf x\|_{pm}&=\left(\frac{1}{n}\sum_{i=1}^n(x_i)^{pm}\right)^{1/(pm)}\nonumber\\
		  &=(\|\mathbf x^m\|_p)^{1/m},\label{eq:id1}
\end{align}
and if $\mathbf x$ has weight $n$
\begin{align}
	\|\mathbf x\|_p&=\frac{1}{\left(\frac{1}{n}\sum_{i=1}^n(x_i)^p\right)^{-1/p}}\nonumber\\
	&=\frac{1}{\|\frac{1}{\mathbf x}\|_{-p}}\label{eq:id2} \ .
\end{align}
From Eq.~\eqref{eq:id1} and Eq.~\eqref{eq:id2} it follows:
\begin{equation}
\|\mathbf x\|_{pm}> \|\mathbf y\|_{pm}\iff \|\mathbf x^m\|_p> \|\mathbf y^m\|_p
\end{equation}
and if $\mathbf x,\mathbf y$ have both weight $n$
\begin{equation}
	\|\mathbf x\|_{p}> \|\mathbf y\|_{p}\iff \left\|\frac {1} {\mathbf x}\right\|_{-p}< \left\|\frac {1}{ \mathbf y}\right\|_{-p} \ .
\end{equation}

Mitra and Ok~\cite{mitra2001majorization} observed that in order to compare inequalities between $p$-norms in the range $(1,\infty)$ it is enough to check the range $(1,r)$ where $r$ depends only on the ratio of $x_1/y_1$ and the dimension of the vectors. The following lemma is a slight generalization of the statement in \cite{mitra2001majorization}:
\begin{lemma}
\label{lem:mitraok}
Let $n>1$, $\mathbf x,\mathbf y\in \mathbb R_+^n$, $y_1>x_1> 0$ and let
\begin{equation}
    r= \frac{\log n}{\log y_1-\log x_1} \nonumber
\end{equation}
Then, $\|\mathbf x\|_p<\|\mathbf y\|_p$ for $\forall p> r$. 
\end{lemma}
\begin{proof}
Consider the following chain of inequalities:
\begin{align}
   \|\mathbf x\|_p^p &\leq (x_1)^p \nonumber\\
                     &\leq \frac{1}{n}(y_1)^p \nonumber\\
                     &\leq \frac{1}{n} \sum_{i=1}^n (y_i)^p \nonumber\\ 
	                 &= \|\mathbf y\|_p^p.
\end{align}

The first inequality holds by our modified definition of $\ell_p$ norm in Equation~\ref{eq:scalednorm_p} and the third inequality holds as all the components of vector $\mathbf{y}$ are non-negative. In addition, the first and third inequality hold independently of the value of $p$. We rewrite the second inequality as:
\begin{equation}
    \left(\frac{y_1}{x_1}\right)^p > n.
\end{equation}
Thus, the second inequality holds if $p> \log n / \log (y_1/x_1)$.
\end{proof}
A similar statement holds for the negative range of $p$ and is as follows:
\begin{corollary}
\label{cor:2} 
Let $n>1$, $\mathbf x,\mathbf y\in \mathbb R_+^n$ with $x_{\min}>y_{\min}>0$ and let
\begin{equation}
    s=\frac{\log n}{\log x_{\min}-\log y_{\min}} \nonumber
\end{equation}
Then, $\|\mathbf x\|_p>\|\mathbf y\|_p$ for $p<-s$. 
\end{corollary}
\begin{proof}
Consider the vectors $\mathbf{1/x}$ and $\mathbf{1/y}$. We have $(1/x)_1=1/x_{\min}$, $(1/y)_1=1/y_{\min}$ and, as a consequence, $(1/y)_1>(1/x)_1>0$.
From Lemma \ref{lem:mitraok}, we have that for $p>s$: $$\left\|\frac{1}{\mathbf x}\right\|_p<\left\|\frac{1}{\mathbf y}\right\|_p\ .$$
Finally, from \eqref{eq:id2}, we have that 
$$\left\|\frac{1}{\mathbf x}\right\|_p<\left\|\frac{1}{\mathbf y}\right\|_p\ \iff \|\mathbf x\|_{-p}> \|\mathbf y\|_{-p}$$
\end{proof}

Using the above results, we are ready to state and prove our main result as Theorem~ \ref{th:main}:
\begin{theorem}
	\label{th:main}
Let $n> 1$, $\mathbf x,\mathbf y\in P_n$, $\mathbf x$ have weight $n$ and choose
\begin{align}
r=\frac{\log n}{\log y_1-\log x_1} \mbox{ and }\bar{r} := \lfloor r+1 \rfloor. \nonumber
\end{align}
If $r>0$ and
\begin{equation}
\label{eq:condition1}
    F_{k,\bar r}(\mathbf x)> F_{k,\bar r}(\mathbf y), \qquad k \in \{\bar{r}, \bar{r}+1, \ldots, n\bar{r}\}, \nonumber
\end{equation}
then $ \mathbf x\in \overline{T( \mathbf y)}$.

If additionally 
\begin{equation}
\label{eq:condition3}
H_1(\mathbf x)>H_1(\mathbf y) \nonumber
\end{equation}
and either 
\begin{enumerate}
    \item the weight of $\mathbf y$ is smaller than $n$, or,
    \item the weight of $\mathbf y$ is $n$, $s>0$ with
\begin{equation}
s=\frac{\log n}{\log x_{\min}-\log y_{\min}} \mbox{ and }\bar{s}:= \lfloor s+1 \rfloor \nonumber  
\end{equation}
and
\begin{equation}
\label{eq:condition2}
    F_{k,1}\left(\frac{1}{\mathbf x^{\bar s}}\right)< F_{k,1}\left(\frac{1}{\mathbf y^{\bar s}}\right), \qquad k \in \{1,2, \ldots, n\} \ , \nonumber
\end{equation}
\end{enumerate}
then $ \mathbf x\in {T( \mathbf y)}$.
\end{theorem}
\begin{proof}
Consider the first implication, that is that $\mathbf x\in \overline{T(\mathbf y)}$ or equivalently from \eqref{eq:condtbarnorm} that $\|\mathbf x\|_p < \|\mathbf y\|_p$ for $p\in(1,\infty)$.
By hypothesis, the following condition holds:
\begin{equation}
\label{eq:condition1P}
    F_{k,\bar r}(\mathbf x)> F_{k,\bar r}(\mathbf y) \qquad k \in \{\bar r, \bar r+1, \ldots, n\bar r\}.
\end{equation}
Then, by Corollary \ref{cor:strictineq} we have that $\|\mathbf x\|_p<\|\mathbf y\|_p$ in the range $p\in(1,\bar r)$. Moreover, since $r>0$, by Lemma \ref{lem:mitraok} we conclude that $\|\mathbf x\|_p<\|\mathbf y\|_p$ also in the range $p\in(r,\infty)$. 

Let us now investigate the second implication. 
To verify that $\mathbf x$ is in $T(\mathbf y)$, it needs to hold additionally that $H_1(\mathbf x) > H_1(\mathbf y)$ and that $\|\mathbf x\|_p > \|\mathbf y\|_p$ for $p\in(-\infty,1)/ \{0\}$ (see \eqref{eq:condtnorm}). The first condition holds by hypothesis, let us verify the second one.

From \eqref{eq:condition1P} we have that $\|\mathbf x\|_p>\|\mathbf y\|_p$ in the range $(0,1)$. It remains to check the range $(-\infty,0)$. We divide the range  into two partially overlapping ranges: $(-\infty,-s)$ and $(-\bar s, 0)$. 

Since $s>0$, we have by Corollary \ref{cor:2} that: $\|\mathbf x\|_p>\|\mathbf y\|_p$ for $p<-s$.

To check the range $(-\bar s,0)$, we have by assumption that:
\begin{equation}
\label{eq:condition2P}
    F_{k,1}\left(\frac{1}{\mathbf x^{\bar s}}\right)< F_{k,1}\left(\frac{1}{\mathbf y^{\bar s}}\right) \qquad k \in \{1,2, \ldots, n\},
\end{equation}
which implies from Corollary \ref{cor:strictineq} that
\begin{equation}
\left\|\frac{1}{\mathbf x^{\bar s}}\right\|_{t} < \left\|\frac{1}{\mathbf y^{\bar s}}\right\|_{t} \qquad t\in(0,1) \label{eq:sbarrange}
\end{equation}
We conclude by showing the equivalence between \eqref{eq:sbarrange} and the desired inequality in the range $(-\bar s,0)$:
\begin{align}
&~~~~~~~~~~\left\|\frac{1}{\mathbf x^{\bar s}}\right\|_{t}<\left\|\frac{1}{\mathbf y^{\bar s}}\right\|_{t} \qquad t\in(0,1)\label{eq:compressedrange}\nonumber\\
&\iff \left\|\frac{1}{\mathbf x}\right\|_{t\bar s}<\left\|\frac{1}{\mathbf y}\right\|_{t\bar s} \qquad t\in(0,1) \nonumber\\
&\iff \left\|\frac{1}{\mathbf x}\right\|_{p}<\left\|\frac{1}{\mathbf y}\right\|_p \qquad p\in(0,\bar s) \nonumber\\
&\iff \|\mathbf x\|_p>\|\mathbf y\|_p \qquad p\in(-\bar s,0) 
 \end{align}
\end{proof}

Thus, Theorem~\ref{th:main} demonstrates a finite set of inequalities that imply the trumping relation for transforming a general state vector $\mathbf x$ to $\mathbf y$ when the allowed free operations are LOCC.

%%%%%%%%%%%%%%%%%%%%%%%%%%%%%%%%%%%%%%%%%%%%%%

\section{A finite sufficient set of conditions for trumping in second law of thermodynamics} \label{s4}
In the previous section, we have established a finite set of sufficient conditions to verify catalytic majorization between state vectors. We now extend our analysis to the domain of quantum thermodynamics, in particular, the so-called second laws \cite{brandao2015second}.

To begin, we define the generalized R\'enyi divergence, generalized free energy, and introduce an embedding map, highlighting its key properties that are pertinent for our analysis in this section.
\begin{definition}
    The generalized R\'enyi divergence of order $p$ (for all $p \in \bR$) between any two states $\rho$ and $\sigma$, such that $supp(\rho) \subseteq supp(\sigma)$, is defined as:
    \begin{align*}
        D_p(\rho||\sigma):= \frac{\mathrm{sign}(p)}{p-1} \log \Tr \left[ \rho^p \sigma^{1-p} \right] 
    \end{align*}
    where for the limiting cases 
    \begin{itemize}
    \item[]{$p=1$: }$D_1(\rho||\sigma):=\Tr \left[ \rho \log \rho - \rho \log \sigma \right]$ is the quantum Kullback-Liebler divergence also known as the von Neumann relative entropy; and 
    \item[]{$p=\infty$: }$D_\infty(\rho||\sigma):=D_{\max}(\rho||\sigma):= \inf \{ \lambda: \rho \leq 2^\lambda \sigma\}$ is known as the $\max$-R\'enyi divergence.
    \end{itemize}
\end{definition}
\begin{definition} Let $d>0$, $v\in\mathbb N^d$ with $v_i>0$ and $\sum_{i=1}^dv_i=N$; the embedding map, denoted by $\cE:P_d \to P_N,\; (N \geq d)$ is defined as:
    \begin{equation}\label{eq:embed_channel}
        \cE(q):=\left\{\underbrace{\frac{q_1}{\nu_1},\dots,\frac{q_1}{\nu_1} }_{\nu_1},\underbrace{\frac{q_2}{\nu_2},\dots,\frac{q_2}{\nu_2} }_{\nu_2},\ldots,\underbrace{\frac{q_d}{\nu_d},\dots,\frac{q_d}{\nu_d} }_{\nu_d}\right\}.
    \end{equation}
\end{definition}
The embedding map $\cE$ is also a valid classical channel (a stochastic matrix) and possesses the following properties:
    \begin{enumerate}
    \item $\cE$ maps the probability distribution $g:=\{ \frac{\nu_i}{N},\frac{\nu_2}{N}, \ldots,\frac{\nu_d}{N} \}$ to the uniform distribution on support size $N$, that is:
    \begin{equation} \label{eq:E_g=unif}
        \cE(g)=\underbrace{\left\{\frac{1}{N},\frac{1}{N},\ldots,\frac{1}{N}\right\}}_{N}:= u_N\; .
    \end{equation}
    \item It preserves equality of the $p$-R\'enyi divergence between any probability distribution ($q$) and $g$ which becomes:  
    \begin{equation} \label{eq:equiv_relentr}
        D_p(q||g) = D_p(\cE(q)||\cE(g)) = D_p(\cE(q)||u_N)\;.
    \end{equation}
    \end{enumerate}
Another important quantity that will be extensively used in our subsequent analysis is the generalized free energy, defined as follows:
\begin{definition} \cite{brandao2015second} \label{def:free_energy}
The generalized free energy of order $p$ of a system with Hamiltonian $H$ of a state $\rho$, is defined as 
$$
\cF_p(\rho, H) := kT[D_p(\rho||\rho_g)-\log Z] = \cF(\rho_g, H) + kT D_p(\rho||\rho_g)
$$
where $\rho_g:=\frac{e^{-\beta H}}{\Tr(e^{-\beta H})}$ is known as the thermal state associated with the system Hamiltonian $H$ and $\beta=1/T$ is the inverse temperature of the surrounding bath. 
\end{definition}

\subsection{For states diagonal in energy eigenbasis}
Here we consider a thermal system evolving according to the Hamiltonian $H$. We examine two states, $\rho$ and $\sigma$, which are simultaneously diagonal in the energy eigenbasis with $q_\rho$ and $q_{\sigma}$ as their respective vector of eigenvalues arranged in descending order of magnitude. Our goal is to provide a finite set of sufficient conditions for transforming $\rho$ to $\sigma$ using a catalyst. We shall use the nomenclature thermo-majorization as an equivalence to the second law of thermodynamics.

It was shown in \cite[Theorem~18]{brandao2015second} that the necessary and sufficient conditions for catalytic thermo-majorization are obtained by comparing the generalized free energies of order $p$ of $\rho$ and $\sigma$ for infinite values of $p$. In particular, if $\cF_p(\rho,H)>\cF_p(\sigma,H)$ for all $p\in (-\infty,\infty)$, catalytic thermo-majorization is possible. This is equivalent (from Definition~\ref{def:free_energy}) to checking $D_p(\rho||\rho_g)>D_p(\sigma||\rho_g)$ for all $p\in (-\infty,\infty)$. 

A technical hurdle in this endeavor is to address the case where the vector of eigenvalues of the thermal state $\rho_g-$ denoted as $g-$ contains irrational entries. We will first consider in Corollary \ref{cor:second_law_rational} a $g$ that contains only rational entries before considering a general $g$ in Theorem \ref{thm:second_law}.

\begin{corollary} \label{cor:second_law_rational}
The following set of finite conditions are sufficient to transform a state $\rho$ with the vector of eigenvalues denoted by $q_\rho$ and is block diagonal in the energy eigenbasis into another block diagonal state $\sigma$ with the vector of eigenvalues denoted by $q_\sigma$ using catalytic thermal operations, given that the vector of eigenvalues of the thermal state $\rho_g$, denoted by $g$, contain all the rational entries: 
\begin{enumerate} \label{eq:second_law}
    \item $F_{k,\bar{r}}(\mathbf x) < F_{k,\bar{r}}(\mathbf y), \qquad k \in \{\bar{r}, \bar{r}+1, \ldots,  N\bar{r}\}$, 
    \item $H_1(\mathbf x)<H_1(\mathbf y)$, and either,
    \item  the weight of $\mathbf y$ is smaller than $n$, or,\\
    the weight of $\mathbf y$ is $n$, and $F_{k,1}\left(\frac{1}{\mathbf x^{\bar s}}\right) > F_{k,1}\left(\frac{1}{\mathbf y^{\bar s}}\right), ~~ k \in \{1,2, \ldots, n\}$, 
    \end{enumerate}
    where $\mathbf{x} \text{ and } \mathbf{y}$ are the probability vectors obtained from $\cE(q_\rho) \text{ and }\cE(q_\sigma)$ and the parameters $r$ and $s$ are chosen to be:
    \[
        r:=\frac{\log N}{\log x_1- \log y_1} \qquad s:=\frac{\log N}{\log y_{\min}- \log x_{\min}}.
    \]
\end{corollary}

\begin{proof}
This corollary follows directly from Theorem~\ref{th:main}. 
For $p\in(1,\bar r)$ and $p \in (\bar r, \infty )$ (from Corollary~\ref{cor:strictineq} and Lemma~\ref{lem:mitraok} respectively) we get (by interchanging the roles of $x$ and $y$):
\begin{equation} \label{eq:xy}
    \lVert x \rVert_p > \lVert y \rVert_p \text{ and thus } H_p(x) < H_p(y).
\end{equation}
Now,
\begin{align*}
    D_p(\rho||\rho_g) \overset{a}{=} D_p(q_{\rho}||g) & \overset{b}{=} D_p(\cE(q_{\rho})||\cE(g))= D_p(x||u_N)\\
    &\overset{c}{=}\log N - H_p(x)\\
    &\overset{d}{>}\log N-H_p(y)\\
    &\overset{e}{=}D_p(y||u_N)=D_p(\cE(q_{\sigma})||\cE(g))\\
        &\overset{f}{=}D_p(q_{\sigma}||g) \overset{g}{=} D_p(\sigma||\rho_g).
\end{align*}
Here (a) and (g) hold since $q_{\rho}, q_{\sigma}, g$ are the vector of eigenvalues of $\rho, \sigma, \rho_g$ respectively and in (b) and (f), we use the properties of the embedding map~\ref{eq:equiv_relentr}, in (c) and (e) we use the relation between $D_p(.||u_N)$ and $H_p(.)$, and in (d) we use Eq.~\ref{eq:xy}.
    
Further, conditions $(2)$ and $(3)$ which are the extension to other values of $p$ (i.e., $p =1, p\in (0,1)$, $p\in (-\bar s, 0)$, and $p\in (-\infty,-\bar s)$) is a direct consequence, leveraging Corollary~\ref{cor:strictineq}, Corollary~\ref{cor:2}, and thus fulfilling the conditions of Theorem~\ref{th:main} (albeit in a reverse direction to ensure the relative R\'enyi entropy aligns with the desired direction for thermo-majorization). Thus, $\rho$ can undergo transformation to $\sigma$ via thermal operation if the conditions outlined in Corollary~\ref{cor:second_law_rational} are met.
\end{proof}

Now, we consider the case where the vector $g$ (vector of eigenvalues of the thermal state of the underlying system) has irrational entries. For this we first approximate $g$ by a vector $g_\epsilon$ that is arranged in descending order of coordinates and is close to $g$ in $\ell_1$ norm. This is stated as the following fact:
\begin{fact}\mbox{\cite[Lemma~15]{brandao2015second}}\label{fact:irrational}
Let $g = \{g_i\}_{i=1}^d$ be a probability distribution of support size $d$, in descending order, and containing irrational entries. Then for any $\epsilon > 0$, there exists a probability distribution $g_\epsilon$ such that:
\begin{enumerate}
\item $\lVert g - g_\epsilon \lVert_1 \leq \epsilon$ ;
\item Each entry of $g_\epsilon$ is a rational number, that is there exists a set of natural numbers $\{\nu_i\}_{i=1}^d$ such that $g_\epsilon = \{\frac{\nu_i}{N}\}_{i=1}^d$,
with $\mathop{\Sigma}\limits_{i=1}^d \nu_i = N$; and
\item There exists a valid classical channel $\cN$ such that $\cN(g) = g_\epsilon$ and for any other distribution $x$, $\lVert x - \cN(x) \rVert_1 \leq O(\sqrt{\epsilon})\; .$
\end{enumerate}
\end{fact}
We give a simple proof of the statement $1$ of the above Fact~\ref{fact:irrational}, directly useful for our purpose in Proposition~\ref{prop:rational_apprx} in the Appendix. Subsequently, given any fixed $\epsilon$, the approximation parameter, we define a density matrix $\rho_{g_\epsilon}$ diagonal in the same basis as $\rho_g$, with the vector of eigenvalues given by $g_\epsilon$. This is now a rational approximation of the actual thermal state and will be used to show catalytic majorization. This approximation parameter, will influence the number of conditions (comparisons of the R\'enyi $p$-entropies) to be checked for catalytic majorization. The formal catalytic majorization theorem for a general thermal state is as follows:     
\begin{theorem} \label{thm:second_law}
Let $\epsilon > 0$ be given and $\rho_{g_{\epsilon}}$ be a density matrix with rational entries that commutes with the thermal state $\rho_g$ and satisfies $\Vert \rho_g-\rho_{g_{\epsilon}} \rVert_1 \leq \epsilon$. The following set of finite conditions are sufficient to transform a state $\rho$ block diagonal in the energy eigenbasis into another block diagonal state $\sigma$ using catalytic thermal operations:
\begin{enumerate}
\item $F_{k,\bar{r}}( \mathbf x) < \frac{F_{k,\bar{r}}( \mathbf y)}{A_r(\epsilon) }, \qquad k \in \{\bar{r}+1, \bar{r}+2, \ldots, N\bar{r}\}$
\item $H_1( \mathbf x) < H_1(\mathbf y)-2 \log \left(1+\frac{\epsilon}{g_{\min}} \right)$, and either,
\item the weight of $\mathbf y$ is smaller than $n$, or,\\
the weight of $\mathbf y$ is $n$, and $F_{k,1}\left(\frac{1}{\mathbf x^{\bar s}}\right) /A_s(\epsilon) > F_{k,1}\left(\frac{1}{\mathbf y^{\bar s}}\right), ~~ k \in \{1,2,\ldots,n\} $.
\end{enumerate}
Here $\mathbf{x} \text{ and } \mathbf{y}$ are the probability vectors obtained from $\cE(q_\rho) \text{ and }\cE(q_\sigma)$, where $q_\rho,q_\sigma$ are the eigenvectors of $\rho,\sigma$ respectively. $A_r(\epsilon)$ and $A_s(\epsilon)$ are some positive real-valued functions, $g_{\min}:=\mathop{\min}\limits_i g_i$, and $\lVert g-g_\epsilon\rVert \leq \epsilon$, for any arbitrary $\epsilon>0$ with $g_\epsilon$ being a vector with rational entries, and $\cE$ is defined in Eq.~\ref{eq:embed_channel}. The functions $A_r(\epsilon)$ and $A_s(\epsilon)$ are such that
\begin{align}
\frac{1}{A_r(\epsilon)}&=\max \left\{ 2^{\left[- \frac{1}{N} \left\{\left(1+ \frac{\epsilon}{g_{\min}}\right)^{2\bar r} \right\} \right]}, 2^{-\frac{2 \epsilon}{N g_{\min}}} \right \}, \nonumber \\
\frac{1}{A_s(\epsilon)}&=2^{-\frac{1}{N}\left\{ \left(1+\frac{\epsilon}{g_{\min}} \right)^{2 (1+ \bar s)}-1\right\}}, \nonumber
\end{align}
and the parameters $r$ and $s$ are chosen as:
\begin{align} %\label{eq:thermo_irrational_r_s}
r&:=\frac{\log N}{\log x_1- \log \left\{ y_1 \left(1+\frac{\epsilon}{g_{\min}} \right)^2\right\}},\nonumber \\
s&:=\frac{\log N}{\log y_{\min}-\log \left\{x_{\min} \left(1+\frac{\epsilon}{g_{\min}} \right)^2\right\}}. \nonumber
\end{align}
\end{theorem}

\begin{proof}
We prove that the conditions stated in the theorem imply $D_p(\rho||\rho_g)>D_p(\sigma||\rho_g)$ for $p\in\mathbb R$ in a piece-wise fashion. First note that for any given $\epsilon>0$, the existence of $\rho_{g_{\epsilon}}$ is shown by Fact~\ref{fact:irrational}. This approximation calls for an additional analysis to relate $D_p(\rho||\rho_g)$ with $D_p(\rho||\rho_{g_\epsilon})$ by proving a continuity argument in Proposition~\ref{prop:continuity_D_alpha}. We now indicate the lemmas and main ideas for each regime, by taking into account this  approximation, essentially leading to $A_r(\epsilon)$ and $A_s(\epsilon)$ in the following analysis: 

For $p=1$, the proof is provided in Lemma~\ref{C2'_lemma2}. 

For the regime $1<p<\bar r$, a detailed proof is presented in Appendix-\ref{sec:thermo_p_r}. Here, we provide a brief outline for completeness. For a sufficiently small positive real-valued function $A_1(\epsilon)$ we have:
\begin{align}\label{eq:proof_sketch}
    &F_{k,\bar{r}}(\mathbf x) <  
        \frac{F_{k,\bar{r}}(\mathbf y)}{A_1(\epsilon)} \quad \text{for
 }k \in \{\bar{r}, \bar{r}+1, \ldots, N\bar{r}\} \nonumber\\
        &\overset{a}{\Rightarrow} f_{\bar{r}}(x,t) < \frac{f_{\bar{r}}(y,t)}{A_1(\epsilon)} \nonumber\\
        &\overset{b}{\Rightarrow} \lVert \mathbf x \rVert_p > \lVert \mathbf y\rVert_p \left( 1+ \frac{\epsilon}{g_{\min}}\right)^{2} \nonumber\\
        &\overset{c}{\Rightarrow} H_p( \mathbf x) < H_p(\mathbf y) -\frac{2p}{p-1} \log \left(1+ \frac{\epsilon}{g_{\min}} \right) \nonumber\\
        &\overset{d}{\Rightarrow} D_p(\rho||\rho_g) > D_p(\sigma||\rho_g) . \nonumber
    \end{align}
where (a), (b), (c), and (d) follows from a series of Lemmas (Lemma \ref{lem:thermal1_high_alpha}, \ref{lem:thermal2_high_alpha}, \ref{lem:thermal3_high_alpha}, and Lemma \ref{lem:thermal4_high_alpha} and \ref{lem:thermal5_high_alpha} respectively) with $\frac{1}{A_1(\epsilon)} \leq 2^{\left[- \frac{1}{N} \left\{\left(1+ \frac{\epsilon}{g_{\min}}\right)^{2\bar r} \right\} \right]}$. 

For the range $p \in (0,1)$, a similar proof is outlined in Appendix-\ref{sec:thermo_p_zero-one}, utilizing a different set of lemmas, specifically Lemmas \ref{lem:thermal2}, \ref{lem:thermal3}, \ref{lem:thermal4}, and \ref{lem:thermal5} and considering $\frac{1}{A_2(\epsilon)} \leq 2^{-\frac{2 \epsilon}{N g_{\min}}}$. 

Since condition 1 is a consequence of the range, $p \in (1, \bar r)$ and $p \in (0,1)$, $\frac{1}{A_r(\epsilon)}$ can be chosen to be as small as $\max \left\{ \frac{1}{A_1(\epsilon)},\frac{1}{A_2(\epsilon)}\right\} $.

We now consider the regime $p \in (-\bar s,0)$. In this range, we have
\begin{align}%\label{eq:proof_sketch_3}
    & F_{k,1}\left(\frac{1}{\mathbf x^{\bar s}}\right) /A_s(\epsilon) > F_{k,1}\left(\frac{1}{\mathbf y^{\bar s}}\right), \quad \text{for }k \in \{1,2,\ldots, N\} \nonumber\\
    &\overset{a}{\Rightarrow} f_1\left(\frac{1}{\mathbf y^{\bar s}},t\right) < \frac{f_1\left(\frac{1}{\mathbf x^{\bar s}},t\right)}{A_s(\epsilon)}, \nonumber\\
    &\overset{b}{\Rightarrow} \left\|\frac{1}{\mathbf x^{\bar s}}\right\|_{\xi} > \left\|\frac{1}{\mathbf y^{\bar s}}\right\|_{\xi} \left( 1 + \frac{\epsilon}{g_{\min}} \right)^{2 \frac{1+\xi \bar s}{\xi}}, \text{ for any } \xi \in (0,1)  \nonumber \\
    &\overset{c}{\Rightarrow}\lVert \mathbf x \rVert_p < \lVert \mathbf y \rVert_p \left( 1 + \frac{\epsilon}{g_{\min}} \right)^{2\frac{1-p}{p}} \nonumber\\
    &\overset{d}{\Rightarrow} H_p(\mathbf x)< H_p(\mathbf y) - 2 \log \left( 1+ \frac{\epsilon}{g_{\min}}\right)  \nonumber\\
    &\overset{e}{\Rightarrow} D_p(\rho||\rho_g) > D_p(\sigma||\rho_g) \;.\nonumber
\end{align}
Here (a) Follows from Lemma \ref{C3_lemma1} with $A_s(\epsilon) \geq 2^{\frac{1}{N}\left\{ \left(1+\frac{\epsilon}{g_{\min}} \right)^{2 (1+ \bar s)}-1\right\}}$, (b) follows from Lemma \ref{C3_lemma2}, (c) follows from Lemma \ref{C3_lemma3}, (d) follows from Lemma \ref{C3_lemma4}, and (e) Lemma \ref{C3_lemma5} and Lemma \ref{C3_lemma6}. 

Note that the direction of inequality (b) is the reverse of that appearing in trumping without the thermal operations. This is because in order to prove thermo-majorization one needs to use the relation between $\lVert \mathbf x \rVert_p, H_p( \mathbf x)$ and $D_p(\rho||\rho_g)$, the definition of the extension map $\cE$, and Eq.~\ref{eq:equiv_relentr}.

Finally, for the regime $p> \overline{r}$ and $p < -\bar s$, the proof is provided in Lemma~\ref{lem:mitraok_thermo}. 

This completes the proof.
\end{proof}

\subsection{For states having coherence}
In the previous subsection, we describe a set of sufficient conditions for the catalytic transformation of states that are diagonal in the energy eigenbasis. However, in general the states under consideration may contain coherence, i.e. off-diagonal elements in the energy eigenbasis of the Hamiltonian. We point to \cite{Streltsov17} for a review on the resource theory of coherence and definitions of transformations.

Bu {\it et. al} \cite{Bu_2016} derived the necessary and sufficient conditions for catalytic transformations under incoherent operations with pure catalysts. More succinctly, for any two incomparable pure states $\ket{\psi}$, $\ket{\phi} \in \mathcal{H}_d$, the necessary and sufficient condition for the existence of a catalyst $\ket{\chi}$ such that $\ket{\psi} \otimes \ket{\chi} \rightarrow \ket{\phi} \otimes \ket{\chi}$ under incoherent operation is
\begin{eqnarray}
    H_p(\psi) & > & H_p(\phi) ~~\forall p \in (-\infty, \infty)/0 \nonumber\\
    \text{and},~~ \Tr(\log[\psi]) & > &\Tr(\log[\phi])
\end{eqnarray}
where $H_p$ is the Renyi entropy defined earlier and $\psi$ (respectively $\phi$) represent the states after removing all off-diagonal coefficients in the relevant basis \cite{Datta_2023}, i.e. $\psi=\sum_i \bra{i}\psi\rangle\langle\psi \ket{i}\ket{i}\bra{i}$ (resp. $\phi=\sum_i \bra{i}\phi\rangle\langle\phi \ket{i}\ket{i}\bra{i}$).

The above conditions are equivalent to the conditions derived in section \ref{s3} and hence we can get the following corollary.
\begin{corollary} \label{thm:second_law_coherence}
The following set of finite conditions are sufficient to transform a pure state $\ket{\psi}$ into another pure state $\ket{\phi}$ using a pure catalyst and incoherent operations:
\begin{enumerate}
\item $F_{k,\bar{r}}(\mathbf x) > F_{k,\bar{r}}(\mathbf y), \qquad k \in \{\bar r, \bar r+1, \ldots, n\bar r\}$  

\item Additionally $H_1(\mathbf x)>H_1(\mathbf y) $ and either 
\item (a) the weight of $\mathbf y$ is smaller than $n$, or,\\
(b) the weight of $\mathbf y$ is $n$, with
$$F_{k,1}\left(\frac{1}{\mathbf x^{\bar s}}\right)< F_{k,1}\left(\frac{1}{\mathbf y^{\bar s}}\right), \qquad k \in \{1,2, \ldots, n\}$$
where $\mathbf x$ and $\mathbf y$ are the eigenvalues of $\psi$ and $\phi$ respectively and $r$ and $s$ are real positive number as defined in Theorem~\ref{th:main}.
\end{enumerate}
\end{corollary}
The proof of the above result can be argued analogously from Theorem~\ref{th:main} where $\mathbf x$ and $\mathbf y$ are the eigenvalues of $\psi$ and $\phi$ respectively. 
Determining a finite set of sufficient conditions for general (mixed) states and mixed catalysis remains an open challenge. Nevertheless, for such general scenario, a finite set of necessary conditions can be derived as outlined in Appendix-\ref{sec:Coherence}. If any of these conditions are not satisfied, it indicates that catalysis is not possible, however the converse is not true.

%%%%%%%%%%%%%%%%%%%%%%%%%%%%%%%%%%%%%%%

\section{Examples}
We now provide some examples of trumping guaranteed by the sets of sufficient conditions derived in the previous sections. 

{\it Trumping bipartite state vectors under LOCC.} First, we present examples of two state vectors $x_{AB}$ and $y_{AB}$, shared between two spatially-separated parties, Alice and Bob. These state vectors are not interconvertible using LOCC since $\mathbf x \nprec \mathbf y$ where $\mathbf x$ and $\mathbf y$ represent their respective vectors of Schmidt coefficient. However, conversion from $x_{AB}$ to $y_{AB}$ becomes feasible when catalysis is allowed.

Consider the vectors, $\mathbf x = ( 0.6100, 0.3045, 0.0435, 0.0420 )$ and $\mathbf y = (0.7315,0.1211,0.1374,0.0100)$. Here $\mathbf x \nprec \mathbf y$. One can also easily reconcile this with the necessary and sufficient conditions mentioned in \cite{bosyk2018lattice} which only resolve the question when the catalyst is $2$-dimensional. However, our findings (presented in Theorem~\ref{th:main}), confirms the existence of a catalyst indicating that trumping is indeed possible (for explicitly checking the conditions see \cite{repo}). In fact, a suitable catalyst for this example can be found in dimension four, $\mathbf c = (0.48, 0.24, 0.16, 0.12)$ implying $\mathbf x \otimes \mathbf c\prec \mathbf y \otimes \mathbf c$. \\

{\it Trumping states diagonal in the energy eigenbasis under thermal operation.} We now present an example from thermodynamics where two states, initially not interconvertible under thermal operations, the states do not meet thermo-majorization conditions, become interconvertible when catalysis is allowed. 

Consider two states $\rho$ and $\sigma$ having vector of eigenvalues $q_{\rho} = (0.936918, 0.0467542, 0.0159775, 0.000350242)$, $q_{\sigma} = (0.862942, 0.129846, 0.00558697, 0.00162474)$ respectively. Here $q_{\sigma} \nprec q_{\rho}$ under thermal operations, where the thermal state $g$ is defined as $g = \frac{1}{Z}(e^{-\beta E_0}, e^{-\beta E_1}, e^{-\beta E_2}, e^{-\beta E_3})$ with $Z= e^{-\beta E_0} + e^{-\beta E_1} + e^{-\beta E_2} + e^{-\beta E_3}$, $\beta = 1.2$ and $E_0=0, E_1 = 1, E_2 =2, E_3 =3$. Although $q_{\sigma}$ and $q_{\rho}$ are not interconvertible under thermal operations, they satisfy the conditions from Theorem~\ref{thm:second_law} when choosing $\rho_\epsilon$ as the maximally mixed state. In fact, in the presence of a catalyst $\mathbf c = (0.48, 0.24, 0.16, 0.12)$, $q_{\rho}$ thermo-majorizes $q_{\sigma}$. 

{\it Trumping general states having coherence.} For states having coherence, consider $|\psi\rangle = \sqrt{0.4}|0\rangle  + \sqrt{0.4}|1\rangle +\sqrt{0.1}|2\rangle+\sqrt{0.1}|3\rangle$ and $|\phi\rangle = \sqrt{0.5}|0\rangle  + \sqrt{0.25}|1\rangle +\sqrt{0.25}|2\rangle$ as presented in \cite{Bu_2016}. Although the vector of eigenvalues of $|\psi\rangle$ and $|\phi\rangle$ do not majorize each other initially, the transformation becomes feasible in the presence of the catalyst, $|\chi\rangle = \sqrt{0.6}|0\rangle  + \sqrt{0.4}|1\rangle$. This transformation is also assured by the finite set of sufficient conditions presented in Corollary~\ref{thm:second_law_coherence}.

The open source software for this numerical exploration can be found in the public repository \cite{repo}.

\section{Discussion}
In this work, we derive a finite set of inequalities that imply trumping. Specifically, we establish a finite set of conditions for trumping of state vectors under LOCC and general state transformation under thermal operations. Our results are also applicable to other restricted subsets of thermal operations known as {\it elementary thermal operations} and {\it Markovian thermal operations}, as the hierarchy of all such thermal processes collapses when catalysis is taken into consideration \cite{son_2023}. Moreover, the analysis presented here is general and we believe that can be non-trivially extended to other concepts, such as Matrix-majorization \cite{Tomamichel24}. Our derivation builds on a set of inequalities relating $\ell_p$ norms to a specific polynomial, as done in \cite{klemes2010symmetric}. We note that, the set of inequalities in \cite{klemes2013more} might be amenable to a similar result. The latter have a special significance in quantum information theory~\cite{jozsa2015symmetric}, as many of the basic entropic quantities have a much more elegant representation when defined over symmetric polynomials.\\

It is interesting to compare the sufficient conditions derived here with other known conditions for the existence of catalyst under certain restrictions on the dimension of the latter. One such condition is given in~\cite{bosyk2018lattice}. There, the authors derive necessary and sufficient conditions for the $4$-dimensional vectors to admit a $2$-dimensional catalyst. It turns out that the set of conditions derived in our work (which works for vectors and catalysts of an arbitrary dimension) is incomparable to their condition. For instance, the example of a pair of $4$-dimensional vectors presented previously satisfies our computable criterion, but fails to satisfy the conditions of~\cite{bosyk2018lattice} thus implying that the catalyst needs to be at least $3$-dimensional. On the other hand, we present the following two vectors which satisfy the criterion of~\cite{bosyk2018lattice} and thus admit a $2$-dimensional catalyst, but in the same time do not satisfy our criterion: $\mathbf x =(0.46519, 0.27313, 0.20361, 0.057807)$, $\mathbf y = (0.46843, 0.2693, 0.20646,  0.05581)$.

It is important to note that the conditions do not shed light on how to find the catalyst. In particular, we do not relate meeting the conditions with a bound on the catalyst dimension. Unfortunately, recent work hints at the computational untractability of this problem \cite{Vjosa}. 

\begin{acknowledgments}
The authors thank Nelly Ng for useful discussions with an earlier version of this project. A.N. thanks Vjosa Blakaj for discussion of her work on level sets of entropy.

S.S. was supported by the Royal Society University Research Fellowship. A.N. acknowledges support from MEXT Quantum Leap Flagship Program (MEXT QLEAP) Grant No. JPMXS0120319794 and ERC Grant Agreement No. 948139. A.N. also acknowledges support by the European Research Council (ERC Grant Agreement No. 948139) and the Excellence Cluster - Matter and Light for Quantum Computing (ML4Q).

The authors declare no competing interests.
\end{acknowledgments}
\bibliography{arXiv_trumping}

%%%%%%%%%%%%%%%%%%%%%%%%%%%%%%%
\appendix
\onecolumngrid
\section{Proof of Theorem~\ref{thm:second_law}: Trumping for thermal processes with thermal states having irrational entries}\label{sec:Thermo_proof}
\subsection{Preliminary results}
\noindent We first introduce the essential ingredients that serve as the foundation for proving the theorem.
\begin{fact}\mbox{\cite[Theorem~2.7.1]{Cover_Thomas}}\label{fact:log_sum}
Consider two sets of positive real numbers $\{ a_i\}_{i=1}^N$ and $\{ b_i\}_{i=1}^N$. The following ineqaulity holds:
$$
\sum\limits_{i=1}^N a_i \log \left( \frac{a_i}{b_i}\right) \geq  \left(\sum\limits_{i=1}^N a_i\right) \log \left( \frac{\sum\limits_{i=1}^N a_i}{\sum\limits_{i=1}^N b_i}\right).
$$
\end{fact}

\noindent We now present the useful definitions and lemmas from \cite{klemes2010symmetric} that connect the scaled $l_p$ norms of vectors to a class of symmetric polynomials used to derive our results. To this end, we start with the following definition:
\begin{definition} \label{def:polynomial}
Consider an integer $r \geq 1$. Denote the $r^{\mathrm{th}}$ degree Taylor polynomial expansion of the exponential function by $P_r$, where $P_r(\nu) = \mathop{\Sigma}\limits_{i=0}^r \frac{\nu^i}{i!}$. For any vector $x \in \bR^N$ and any real number $t \in \bR$ define
\begin{equation} \label{eqn:f_polynomial}
f_r(x,t):= \prod\limits_{i=1}^N P_r(x_i t) =  \prod\limits_{i=1}^N \left( \sum\limits_{j=0}^r \frac{{(x_it)}^j}{j!} \right).
\end{equation}
Further, for any integer $k \geq 1$ we define $F_{k,r}(x)$ as the coefficient of $t^k$ in the expansion of $f_r(x,t)$. Equivalently, we can express $f_r(x,t)$ as:
\begin{equation} \label{eqn:F_polynomial}
    f_r(x,t):= \sum\limits_{k=1}^r F_{k,r}(x)t^k.
\end{equation}
\end{definition}
\begin{fact}\mbox{\cite[Theorem~1]{klemes2010symmetric}}\label{fact:norm_connects_F}
Consider two probability vectors $x,\;y$ of support size $N$ and take a fixed integer $r \geq 1$. Then we have the following characterization of the scaled $l_p$ with respect to the polynomial coefficients $F_{k,r}$ of the exponential function, for all integers $k$ such that $r \leq k \leq Nr$.

If $F_{k,r}(x) \leq F_{k,r}(y)$ then 
\begin{eqnarray*}
\lVert x \rVert_p
& \leq & \lVert y \rVert_p \quad \text{ for }p\in[0,1],\\
\lVert x \rVert_p 
& \geq &
\lVert y \rVert_p \quad \text{ for }p\in[1,r+1]\; .
\end{eqnarray*}
\end{fact}
\noindent We also need the following relations of the so-called Mellin transforms \cite{klemes2010symmetric}:
\begin{fact} \label{fact:integral}
Define the integral transform of the logarithm of $P_r$ (from Definition~\ref{def:polynomial}) as:
\begin{eqnarray*}
I_r(p)
& := &
\int_0^\infty \log \left( \sum\limits_{i=0}^r \frac{\nu^i}{i!} \right) \nu^{-p} \frac{d\nu}{\nu} \textrm{ for }p\in(0,1) \text{ and }\\
J_r(p)
& := &
\int_0^\infty \left( \nu - \log \left( \sum\limits_{i=0}^r \frac{\nu^i}{i!} \right) \right) \nu^{-p} \frac{d\nu}{\nu} \textrm{ for }p\in(1,r+1)\; .
\end{eqnarray*}
Then, if we let $t=\nu/a$ for $a\geq 0$, the following relations hold:
\begin{eqnarray*}
    \frac{1}{I_r(p)}\int_0^\infty \log \left( \sum\limits_{i=0}^r \frac{(at)^i}{i!} \right) t^{-p} \frac{dt}{t}
    &  =  &
    a^p \quad \textrm{ for } p\in(0,1)\\
    \frac{1}{J_r(p)}\int_0^\infty \left( at- \log \left(\sum\limits_{i=0}^r \frac{(at)^i}{i!} \right) \right) t^{-p} \frac{dt}{t}
    &  =  &
    a^p \quad \textrm{ for } p\in(1,r+1).
\end{eqnarray*}
\end{fact}
\noindent We now prove the statement $1$ of Fact~\ref{fact:irrational} to analyze the case when the thermal distribution has irrational entries as the following proposition:\\

\begin{proposition} \label{prop:rational_apprx}
    Consider the spectral decomposition of the thermal state $\rho_g$ in the energy eigenbasis $\{\ket{e_i}\}_{i=1}^d$ as:
    $$
        \rho_g=\mathop{\Sigma}\limits_{i=1}^d g_i \ket{e_i}\bra{e_i}
    $$
    with some of the eigenvalues $g_i$ being irrational. Then there exists a state $\rho_{g_{\epsilon}}$ with all eigenvalues as rational numbers of form $\{\frac{d_i}{N}\}_{i=1}^N$, for $N$ large enough, $\rho_{g_{\epsilon}}$ is diagonal in the energy eigenbasis and $$
        \lVert \rho_g - \rho_{g_\epsilon} \rVert_1 \leq \epsilon
    $$
    for any given $\epsilon>0$.
\end{proposition}
\begin{proof}
    The above proposition is simply an extension of the fact that any irrational number can be approximated by a rational number to any desired accuracy.\\
    More formally, let $g_\epsilon$ denote the vector of eigenvalues of $\rho_{g_\epsilon}$ and $N, \epsilon$ are given. Suppose for some $1 \leq i \leq d$, $g_i$ is irrational. Then by Archimedian property of real numbers, there exists an integer $N'$ such that 
    $N'>\max{[\max_i{[g_i]}, \frac{N}{\epsilon}]}$. Now we define $d_i:=\lfloor g_i N' \rfloor$ for all $1 \leq i \leq d$. \\
    Since $g$ is a probability vector, therefore $\mathop{\Sigma}\limits_{i=1}^d d_i=N'$.\\
    Finally, define the rational approximation of $g_i$ as: 
    $
    (g_\epsilon)_i:=\frac{d_i}{N'}\; .
    $
    Thus, $(g_\epsilon)_i$ satisfy:
    \begin{align*}
        |g_i-(g_\epsilon)_i| = \left|g_i - \frac{d_i}{N'} \right|
        = \left| \frac{N' g_i - d_i}{N'}\right|
        & \leq \frac{\epsilon}{N'} \; .
    \end{align*}
    Note that if for any $1\leq i \leq d, \text{ if } g_i$ is rational then $g_i=(g_\epsilon)_i=\frac{d_i}{N}$. We thus obtain the resultant density matrix
    $$
    \rho_{g_\epsilon}:=\sum\limits_{i=1}^d \frac{d_i}{N'} \ket{e_i} \bra{e_i}\; .
    $$ 
    This satisfy the property that:
    \begin{align*}
        \lVert \rho_g - \rho_{g_\epsilon}\rVert_1 =  \sum\limits_{i=1}^d \left| g_i-\frac{d_i}{N'}\right|
        \leq N \frac{\epsilon}{N'} \leq \epsilon . 
    \end{align*}
\end{proof}
\noindent Below we prove a continuity statement for the $p$-R\'enyi divergence.

\begin{proposition}\label{prop:continuity_D_alpha}
Let $p>0$, and any given $\epsilon > 0$, and two probability distributions $x$ and $g$. Let $g_{\min}:=\min_i g_i$ and $g_\epsilon$ be another probability distributions such that $\lVert g-g_\epsilon\rVert_1 \leq \epsilon$, then:
\begin{eqnarray} \label{eq:cont_1}
\left| D_p(x||g_\epsilon)-D_p(x||g) \right| 
& \leq & \max\left\{1, \frac{p}{|p-1|} \right\} \log \left( 1+\frac{\epsilon}{g_{\min}}\right), \text{ and }
\end{eqnarray}
\begin{eqnarray} \label{eq:cont_p1}
\left| D_1(x||g_\epsilon)-D_1(x||g) \right| 
& \leq & \log \left( 1+\frac{\epsilon}{g_{\min}}\right) .
\end{eqnarray}
\end{proposition}
\begin{proof}
The above proposition can be proven as follows:
\begin{eqnarray*}
    |D_p(x||g_\epsilon)-D_p(x||g)|
    & = &
     \frac{1}{|p-1|} \bigg|\left\{ \log \left( \mathop{\Sigma}\limits_{i=1}^N x_i^p (g_\epsilon)_i^{1-p}\right)- \log \left( \mathop{\Sigma}\limits_{i=1}^N x_i^p g_i^{1-p} \right) \right\}\bigg|\\
    & = &
    \frac{1}{|p-1|\left(\mathop{\Sigma}\limits_{i=1}^N x_i^p {(g_\epsilon)_i}^{1-p}\right)}  \left| \left(\mathop{\Sigma}\limits_{i=1}^N x_i^p {(g_\epsilon)_i}^{1-p}\right) \log \left( \frac{\mathop{\Sigma}\limits_{i=1}^N x_i^p {(g_\epsilon)_i}^{1-p}}{\mathop{\Sigma}\limits_{i=1}^N x_i^p {g_i}^{1-p}} \right) \right|\\
    & \overset{a}{\leq} &
    \frac{1}{|p-1|\left(\mathop{\Sigma}\limits_{i=1}^N x_i^p {(g_\epsilon)_i}^{1-p}\right)}  \mathop{\Sigma}\limits_{i=1}^N x_i^p {(g_\epsilon)_i}^{1-p} \left| \log {\left( \frac{(g_\epsilon)_i}{g_i}\right)}^{1-p} \right|\\
    & \leq &
    \frac{1}{\left(\mathop{\Sigma}\limits_{i=1}^N x_i^p {(g_\epsilon)_i}^{1-p}\right)} \left(\mathop{\Sigma}\limits_{i=1}^N x_i^p {(g_\epsilon)_i}^{1-p} \right)  \left[ \max_i \left|\log \left( \frac{(g_\epsilon)_i)}{g_i}\right) \right| \right]\\
    & = &
    \max \limits_i \left| \log \left( \frac{(g_\epsilon)_i}{g_i}\right) \right|\\
    & \overset{b}{\leq} &
    \log \left( 1+ \frac{\epsilon}{g_{\min}} \right),
\end{eqnarray*}
where (a) follows from the log-sum inequality, fact~\ref{fact:log_sum}, 
 and (b) follows from the following:
\begin{eqnarray} \label{eq:l1_norm}
    \epsilon \geq \lVert g-g_\epsilon \rVert_1 
    & = &
    \mathop{\Sigma}\limits_{i=1}^N g_i \left|\frac{(g_\epsilon)_i}{g_i}-1\right| \notag\\
    & \geq &
    g_{\min} \mathop{\Sigma}\limits_{i=1}^N \left|\frac{(g_\epsilon)_i}{g_i}-1\right| \notag\\
    & \geq &
    g_{\min} \max_i \left|\frac{(g_\epsilon)_i}{g_i}-1\right| \notag\\
    & \Rightarrow & \log \left( \frac{(g_\epsilon)_i}{g_i}\right) \leq \log \left( 1+ \frac{\epsilon}{g_{\min}}\right); \textrm{ for all $i \in $ }\{ 1,2 \ldots, N\}.
\end{eqnarray}

We now give a continuity statement for the case of $p=1$ (i.e., for KL-divergence). The simplicity comes from the definition of $D_1(x||g_\epsilon):= \mathop{\Sigma}\limits_i x_i \log \frac{x_i}{(g_\epsilon)_i}$ which leads to:
\begin{align}
    |D_1(x||g)-D_1(x||g_{\epsilon})| &= \mathop{\Sigma}\limits_i x_i \left|\log \left( \frac{(g_\epsilon)_i}{g_i} \right) \right| \notag\\
    &\leq \max_i \left|\log \left( \frac{(g_\epsilon)_i}{g_i} \right) \right| \mathop{\Sigma}\limits_i x_i \notag\\
    &=\max_i \left|\log \left( \frac{(g_\epsilon)_i}{g_i} \right) \right| \notag\\
    & \overset{i}{\leq}  \log \left( 1+ \frac{\epsilon}{g_{\min}}\right)
\end{align}
where (i) follows from Eq.~\ref{eq:l1_norm}. 
This proves Eq.~\ref{eq:cont_p1}.
\end{proof}

\noindent In the following lemmas, we demonstrate how conditions based on symmetric polynomials lead to trumping by explicitly examining the relation between $D_p(\rho||\rho_g) $ and $ D_p(\sigma||\rho_g)$ across the full range of $p \in \mathbb{R}$. Throughout, we consider $x,\;y$ as two probability vectors of support size $N$, that is, $\mathop{\Sigma}\limits_{i=1}^N x_i=\mathop{\Sigma}\limits_{i=1}^N y_i=1$ and without loss of generality, the logarithm is taken with base $2$. 

%%%%%%%%%%%%%%%%%%%%%%%%%%%%%%%%%%%%%%%%%%%%%%%%%%

\subsection{Characterization in terms of relative entropy, for $p=1$:} \label{sec:thermo_p1}
\noindent For $p=1$, following two lemma guarantee the conditions of trumping.
\begin{lemma}\label{C2'_lemma2}
    For any given $\epsilon>0$, if $H_1(x) < H_1(y)-2 \log \left(1+\frac{\epsilon}{g_{\min}} \right)$, then
    \[
    D_1(\rho||\rho_{g_\epsilon}) > D_1(\sigma||\rho_{g_\epsilon}) + 2 \log \left(1+\frac{\epsilon}{g_{\min}} \right) \;.
    \]
    \end{lemma}
    \begin{proof}
        The proof follows easily by using the following relation between entropy and the relative entropy:
        \[
            D_1(x||g_\epsilon)=\log N -H_1(x)\;.
        \]
        Using the hypothesis of the lemma and above relation completes the proof.
    \end{proof}
\begin{lemma}\label{C2''_lemma2}
    For any given $\epsilon>0$, if $D_1(\rho||\rho_{g_\epsilon}) > D_1(\sigma||\rho_{g_\epsilon}) + 2 \log \left(1+\frac{\epsilon}{g_{\min}} \right)$, then
    \[
        D_1(\rho||\rho_g)> D_1(\sigma||\rho_g).
    \]
    \begin{proof}
        The continuity of relative entropy from Eq.~\ref{eq:cont_p1} together with the hypothesis of the lemma, gives
        \begin{align*}
            D_1(\rho||\rho_g) &\overset{a}{>} D_1(\rho||\rho_{g_\epsilon})-\log \left( 1+ \frac{\epsilon}{g_{\min}}\right)\\
            & > D_1(\sigma||\rho_{g_\epsilon})+\log \left( 1+ \frac{\epsilon}{g_{\min}}\right)\\
            & \overset{b}{>} D_1(\sigma||\rho_{g})
        \end{align*}
        where (a) and (b) follows from Eq.~\ref{eq:cont_p1}.
        %This proves the lemma.
    \end{proof}
\end{lemma}

%%%%%%%%%%%%%%%%%%%%%%%%%%%%%%%%%%%%%%%%%%%%%%%%%%%
\subsection{Characterization in terms of $p$ R\'enyi divergence, for $p\in(1,\bar r)$:}\label{sec:thermo_p_r}
\noindent For $p\in(1,\bar r)$, in order to show $F_{k,\bar r}(x) < \frac{F_{k, \bar r}(y)}{A_1(\epsilon)}$ implies trumping (for some suitably chosen $A_1(\epsilon)$), we prove following series of Lemmas.
\begin{lemma}\label{lem:thermal1_high_alpha}
If $F_{k,\bar r}(x) < \frac{F_{k,\bar r}(y)}{A_1(\epsilon)}$ for some real valued positive function $A_1(\epsilon)$ which is independent of $k$ (with $\epsilon>0$) then
$$
f_{\bar r}(x,t) < \frac{f_{\bar r}(y,t)}{A_1(\epsilon)} .
$$
\end{lemma}
\begin{proof}
    By definition~\ref{def:polynomial} of $f_{\bar r}(x,t)$ and $F_{k,\bar r}(x)$ \cite{klemes2010symmetric}:
    \begin{align*}
        f_{\bar r}(x,t) &= \mathop{\Sigma}\limits_{k}F_{k,\bar r}(x)t^k \\
        &\overset{a}{<} \frac{1}{A_1(\epsilon)}\mathop{\Sigma}\limits_{k}F_{k, \bar r}(y)t^k\\
        & = \frac{f_{\bar r}(y,t)}{A_1(\epsilon)}.
    \end{align*}
    Here (a) follows from the hypothesis of the lemma.
\end{proof}

\begin{lemma}\label{lem:thermal2_high_alpha}
If $f_{\bar r}(x,t) < \frac{f_{\bar r}(y,t)}{A_1(\epsilon)}$ for any $\epsilon>0$ and $A_1(\epsilon)$ is as defined in Lemma~\ref{lem:thermal1_high_alpha} and in addition satisfies $A_1(\epsilon) \geq
2^{\left[ \frac{1}{N} \left\{\left(1+ \frac{\epsilon}{g_{\min}}\right)^{2 \bar r} \right\} \right]}$, then
$$
\lVert x \rVert_p > \lVert y \rVert_p \left( 1 + \frac{\epsilon}{g_{\min}} \right)^{2} .
$$
\end{lemma}
\begin{proof}
    Using Fact~\ref{fact:integral}, we have the following:
    \begin{eqnarray*}
    \mathop{\Sigma}\limits_{i=1}^N
\left[ y_i t- \log \left( \mathop{\Sigma}_{j=1}^{\bar r} \frac{{(y_i t)}^j}{j!}\right) \right]
& = &
    \mathop{\Sigma}\limits_{i=1}^N y_i t- \log f_{\bar r}(y,t) \\
    & \overset{a}{<} &
     \mathop{\Sigma}\limits_{i=1}^N y_i t- \log f_{\bar r}(x,t) - \log A_1(\epsilon).
     \end{eqnarray*}
Here (a) follows from the hypothesis of the lemma. Now multiplying both sides of the above Eq. by $\frac{t^{-p-1}}{J_{\bar r}(p)}$, rearranging the terms, integrating with respect to $t \in (0, \infty)$ and using Fact~\ref{fact:integral}, we get:
     \begin{eqnarray*}
N \lVert x \rVert_p^p
        &  >  &
        N \lVert y \rVert_p^p + \frac{\log A_1(\epsilon)}{J_{\bar r}(p)} \int_0^\infty t^{-p-1} dt\\
        & \overset{a}{>} &
        N \lVert y \rVert_p^p + \frac{\log A_1(\epsilon)}{J_{\bar r}(p)} \int_{k}^\infty t^{-p-1} dt\\
        & = &
        N \lVert y \rVert_p^p + N \log A_1(\epsilon)\\
         \Rightarrow \lVert x \rVert_p^p
         & > &
         \lVert y \rVert_p^p \times \left[ 1+\frac{\log A_1(\epsilon)}{ \lVert y \rVert_p^p} \right]\\ 
         \Rightarrow \lVert x \rVert_p
         & \overset{b}{>} &
         \lVert y \rVert_p { \left[ 1+ \frac{\epsilon}{g_{\min}}\right] }^{2},\;
    \end{eqnarray*}
    where in (a) we set the lower limit of the integral as $k={(N p J_{\bar r}(p))}^{-\frac{1}{p}}>0$ ensuring that the integral evaluates to $NJ_{\bar r}(p)$ and (b) holds for the choice of $A_1(\epsilon)$ as:
    \begin{align*}
     A_1(\epsilon) &\geq 2^{\left[ \frac{1}{N} \left\{\left(1+ \frac{\epsilon}{g_{\min}}\right)^{2 \bar r} \right\} \right]} \\
     & \overset{a}{\geq} 2^{ \lVert y \rVert_p^p \left\{ {\left(1+ \frac{\epsilon}{g_{\min}} \right)}^{2 \bar r)} \right\}}\\
     & \overset{b}{\geq} 2^{ \lVert y \rVert_p^p \left\{ {\left(1+ \frac{\epsilon}{g_{\min}} \right)}^{2p}-1 \right\}}.
    \end{align*}
    In the above analysis (a) holds because $\lVert y \rVert_p^p= \frac{\Sigma_{i=1}^n y_i^p}{N} \leq \frac{\Sigma_{i=1}^N y_i}{N}=\frac{1}{N}$, since $y$ is a probability vector and (b) follows as $p \in (1,\bar r)$. \\
    This completes the proof of the lemma.
\end{proof}

\begin{lemma}\label{lem:thermal3_high_alpha}
If $\lVert x \rVert_p > \lVert y \rVert_p {\left( 1+ \frac{\epsilon}{g_{\min}}\right)}^{2}$ for any given $\epsilon>0$, then:
    $$
    H_p(x)< H_p(y) - \frac{2p}{p-1} \log \left( 1+ \frac{\epsilon}{g_{\min}}\right) .
    $$
\end{lemma}
\begin{proof}
    From the definition of Renyi entropy (Eq. 3, 4, 5) we have:
    \begin{eqnarray*}
        H_p(x)
        &  :=  &
        -\frac{1}{p-1}\log \left( N \times \frac{\mathop{\Sigma}\limits_{i=1}^N x_i^p}{N}\right)\\
        &  \overset{a}{<}  &
        -\frac{\log N}{p-1} - \frac{p}{p-1} \log \lVert y \rVert_p - \frac{2p}{p-1} \log \left( 1+\frac{\epsilon}{g_{\min}}\right)\\
        &  =  &
        H_p(y) - \frac{2p}{p-1} \log \left( 1+\frac{\epsilon}{g_{\min}}\right),
    \end{eqnarray*}
    where (a) follows from the hypothesis of the lemma. %This completes the proof of the lemma.
\end{proof}

\begin{lemma}\label{lem:thermal4_high_alpha}
For any given $\epsilon>0$, if $H_p(x) < H_p(y) - \frac{2p}{p-1} \log \left( 1+ \frac{\epsilon}{g_{\min}}\right) $ then:
\begin{eqnarray} \label{eq:termal4_2}
D_p(\rho||\rho_{g_\epsilon}) & \geq & D_p(\sigma||\rho_{g_\epsilon})+\frac{2p}{p-1} \log \left( 1+ \frac{\epsilon}{g_{\min}}\right) .
\end{eqnarray}
\end{lemma}
\begin{proof}
The proof essentially follows from the relation between the relative entropy of a given distribution with uniform distribution and the Shannon entropy of a given state, i.e., $D_p(x||u_N)=\log N - H_p(x)$ and similarly $D_p(y||u_N)=\log N - H_p(y)$. Since,
     \begin{eqnarray*}
        D_p(\rho||\rho_{g_\epsilon}) 
        & = & 
        D_p(x||u_N) =\log N - H_p(x)\\
        &  \overset{a}{>}  &
        \log N - H_p(y) + \frac{2p}{p-1} \log \left( 1+ \frac{\epsilon}{g_{\min}}\right)\\
        &  =  &
        D_p(y||u_N) + \frac{2p}{p-1} \log \left( 1+ \frac{\epsilon}{g_{\min}}\right)\\
        & = &
        D_p(\sigma||\rho_{g_\epsilon}) + \frac{2p}{p-1} \log \left( 1+ \frac{\epsilon}{g_{\min}}\right) .
    \end{eqnarray*}
    Here (a) follows from the hypothesis of the lemma.
\end{proof}

\begin{lemma}\label{lem:thermal5_high_alpha}
    For any given $\epsilon>0$, if $D_p(\rho||\rho_{g_\epsilon}) \geq D_p(\sigma||\rho_{g_\epsilon})+\frac{2p}{p-1} \log \left( 1+ \frac{\epsilon}{g_{\min}}\right)$ then:
$$
D_p(\rho||\rho_g) > D_p(\sigma||\rho_g).
$$
\end{lemma}
\begin{proof}
    \begin{eqnarray*}
    D_p(\rho||\rho_g)
    & \overset{a}{>} &
    D_p(\rho||\rho_{g_\epsilon}) - \frac{p}{p-1}\log \left( 1+ \frac{\epsilon}{g_{\min}} \right)\\
    & \overset{b}{>} &
    D_p(\sigma||\rho_{g_\epsilon})+\frac{2p}{p-1} \log \left( 1+ \frac{\epsilon}{g_{\min}}\right) -  \frac{p}{p-1}\log \left( 1+ \frac{\epsilon}{g_{\min}} \right)\\
    & \overset{c}{>} &
    D_p(\sigma||\rho_{g})- \frac{p}{p-1}\log \left( 1+ \frac{\epsilon}{g_{\min}} \right) +\frac{2p}{p-1} \log \left( 1+ \frac{\epsilon}{g_{\min}}\right) - \frac{p}{p-1}\log \left( 1+ \frac{\epsilon}{g_{\min}} \right)\\
    & = &
    D_p(\sigma||\rho_{g})
    \end{eqnarray*}
    where (a) follows from Proposition \ref{prop:continuity_D_alpha}, (b) follows from the hypothesis of the Lemma and (c) follows from Proposition \ref{prop:continuity_D_alpha}.
\end{proof}
\noindent Thus, $F_{k,\bar r}(x) < \frac{F_{k,\bar r}(y)}{A_1(\epsilon)}$ along with the choice of $A_1(\epsilon) =
2^{\left[ \frac{1}{N} \left\{\left(1+ \frac{\epsilon}{g_{\min}}\right)^{2 \bar r} \right\} \right]}$ is sufficient to guarantee trumping.

\subsection{Characterization in terms of $p$ R\'enyi divergence, for $p \in (0,1)$:}\label{sec:thermo_p_zero-one}

\noindent Since the relation between scaled $l_p$ norms of two vectors differ in the regime where $p \in (0,1)$ compared to when $p \in (1,\bar r)$, some of the lemmas presented above have different bounds and quantifiers. Here we prove the lemmas for the region $p \in (0,1)$ for clarity and completeness.
\begin{lemma}\label{lem:thermal2}
If $f_{\bar r}(x,t) < \frac{f_{\bar r}(y,t)}{A_2(\epsilon)}$ for any $\epsilon>0$ and $A_2(\epsilon)$ is as defined in Lemma~\ref{lem:thermal1_high_alpha} and in addition satisfies 
$A_2(\epsilon) \geq 2^{\frac{2 \epsilon}{N g_{\min}}}$, then
\begin{align*}
\lVert x \rVert_p <
\begin{cases}
 \lVert y \rVert_p \left( 1 + \frac{\epsilon}{g_{\min}} \right)^{-2 \left(\frac{1-p}{p} \right)} & \text{ for }p \in [0,\frac{1}{2}]\\
 \lVert y \rVert_p \left( 1 + \frac{\epsilon}{g_{\min}} \right)^{-2} & \text{ for }p \in (\frac{1}{2},1]
\end{cases}
\end{align*}
\end{lemma}
\begin{proof}
    Using Fact~\ref{fact:norm_connects_F}, we have the following:
    \begin{eqnarray} \label{eq:norm_lb_pos}
        N \lVert x \rVert_p^p
        &  =  &
        \frac{1}{I_{\bar r}(p)} \int_0^\infty t^{-p-1} \log f_{\bar r}(x,t) dt \notag\\
        &  \overset{a}{<} &
        \frac{1}{I_{\bar r}(p)} \int_0^\infty t^{-p-1} \log f_{\bar r}(y,t) dt - 
        \log A_2(\epsilon) \int_0^\infty \frac{t^{-p-1}}{I_{\bar r}(p)} dt \notag\\
        &  \overset{b}{<} &
        N \lVert y \rVert_p^p - \log A_2(\epsilon) \int_{k}^\infty \frac{t^{-p-1}}{I_{\bar r}(p)} dt \notag\\
        & = &
        N \lVert y \rVert_p^p \left\{ 1- N\eta(p)\log A_2(\epsilon)\right\}. 
    \end{eqnarray}
    where (a) follows from the hypothesis of the lemma, while inequality (b) arises from restricting the integration range of the second term to a modified lower bound $k={(\eta(p)p N^2 \lVert y \rVert_p^p I_{\bar r}(p))}^{-1/p}$ instead of $0$. This ensures that the definite integral in the second term evaluates to $N^2 \lVert y \rVert_p^p  \eta(p)$ with the chosen function $\eta(p)=\max\{p, (1-p) \}$ depending on the range of $p$.\\ 
    Now, we choose $A_2(\epsilon) \geq 2^{\frac{2 \epsilon}{Ng_{\min}}}$ (implying that $\log A_2(\epsilon)>0$), which further leads to:
    \begin{align}\label{eq:simpleA}
        \eta(p)\log A_2(\epsilon) &\geq \frac{2 \eta(p) \epsilon}{N g_{\min}} \notag\\
        & = \frac{1}{N}\left[ 1- \left( 1- \frac{2 \eta(p) \epsilon}{ g_{\min}} \right) \right] \notag\\
        & \overset{i}{\geq} \frac{1}{N}\left[ 1- \left( 1+ \frac{\epsilon}{ g_{\min}} \right)^{-2 \eta(p)} \right]\;,
    \end{align}
    where (i) holds since $(1+x)^{-n} \geq 1-nx$, for $x \in [0,1]$.\\
    \begin{itemize}
    \item[]{Case 1: For $p \in (1/2,1]$}, we have $\eta(p)=p$ and then substituting Eq.~\ref{eq:simpleA} in Eq.~\ref{eq:norm_lb_pos} gives: 
    \[
    \lVert x \rVert_p < \lVert y \rVert_p \left( 1 + \frac{\epsilon}{g_{\min}} \right)^{-2}, \text{ for }p \in (1/2,1]\;.
    \]
    \item[]{Case 2: For $p \in [0,1/2)$},
    we have $\eta(p)=1-p$ and again :
    \[
        \log A_2(\epsilon) \geq \frac{1}{N}\left[ 1- \left( 1+ \frac{\epsilon}{ g_{\min}} \right)^{-2(1-p)} \right].
    \]
    %And hence  implies 
    Substituting Eq.~\ref{eq:simpleA} in Eq.~\ref{eq:norm_lb_pos} gives 
    \[
    \lVert x \rVert_p < \lVert y \rVert_p \left( 1 + \frac{\epsilon}{g_{\min}} \right)^{-2 \left( \frac{1-p}{p}\right)} \text { for }p \in [0,1/2).
    \]
    \end{itemize}
    This completes the proof of the lemma.
\end{proof}
\begin{lemma}\label{lem:thermal3}
If $\lVert x \rVert_p < \lVert y \rVert_p {\left( 1+ \frac{\epsilon}{g_{\min}}\right)}^{-2\gamma \;}$ where $\gamma := \max\left\{ \left( \frac{1-p}{p}\right), 1\right\}$ for any given $\epsilon>0$, then:
    $$
    H_p(x) < H_p(y)  - 2 \frac{\gamma \;p}{1-p} \log \left( 1+ \frac{\epsilon}{g_{\min}}\right) .
    $$
\end{lemma}
\begin{proof}
    From the definition of Renyi entropy (Eq. 3, 4, 5)  we have:
    \begin{eqnarray*}
        H_p(x)
        &  :=  &
        \frac{1}{1-p}\log \left( N \times \frac{\mathop{\Sigma}\limits_{i=1}^N x_i^p}{N}\right)\\
        &  \overset{a}{<}  &
        \frac{\log N}{1-p} + \frac{p}{1-p} \log \lVert y \rVert_p - 2  \frac{\gamma \;p}{1-p}\log \left( 1+\frac{\epsilon}{g_{\min}}\right)\\
        &  =  &
        H_p(y) - 2  \frac{\gamma \;p}{1-p} \log \left( 1+\frac{\epsilon}{g_{\min}}\right)
    \end{eqnarray*}
    where (a) follows from the hypothesis of the lemma. The lemma thus implies:
    \begin{align*}
        H_p(x) <
        \begin{cases}
            H_p(y)- 2\log \left( 1+ \frac{\epsilon}{g_{\min}}\right) & \text{ for } p \in [0,1/2)\\
            H_p(y)- \frac{2p}{1-p}\log \left( 1+ \frac{\epsilon}{g_{\min}}\right) & \text{ for } p \in [1/2,1)
        \end{cases}
    \end{align*}
    This completes the proof of the lemma.
\end{proof}
\begin{lemma}\label{lem:thermal4}
For any given $\epsilon>0$, if $H_p(x)<H_p(y)-2 \frac{\gamma \;p}{1-p} \log \left( 1+ \frac{\epsilon}{g_{\min}}\right) $, for $\gamma := \max\left\{ \left( \frac{1-p}{p}\right), 1\right\}$ then:
$$
D_p(\rho||\rho_{g_\epsilon}) \geq D_p(\sigma||\rho_{g_\epsilon})+ 2  \frac{\gamma \;p}{1-p} \log \left( 1+ \frac{\epsilon}{g_{\min}}\right) .
$$
\end{lemma}
\begin{proof}
    The proof essentially follows from the relation between the relative entropy of a given state with maximally mixed state and the von Neumann entropy of a given state, very similar to that of Lemma~\ref{lem:thermal4_high_alpha} (with the factor $p-1$ being replaced by $1-p$ in the denominator of the second term on the right hand side of the desired inequality).
\end{proof}
\begin{lemma}\label{lem:thermal5}
For any given $\epsilon>0$, if $D_p(\rho||\rho_{g_\epsilon}) \geq D_p(\sigma||\rho_{g_\epsilon}+ 2 \frac{\gamma \;p}{1-p} \log \left( 1+ \frac{\epsilon}{g_{\min}}\right)$ for $\gamma := \max\left\{ \left( \frac{1-p}{p}\right), 1\right\}$ then:
$$
D_p(\rho||\rho_g) > D_p(\sigma||\rho_g) .
$$
\end{lemma}
\begin{proof}
The essential idea is to use the continuity of $D_p(\rho||\rho_g)$ in  the second argument, which follows from Proposition~\ref{prop:continuity_D_alpha}.\\
We do the analysis under the following two cases: 
\begin{itemize}
\item[]{Case 1: For $p \in [0,1/2)$} range, we have $\frac{\gamma\;p}{1-p}=1$. This leads to:
\begin{eqnarray*}
    D_p(\rho||\rho_g)
    &  \overset{i}{>}  &
    D_p(\rho||\rho_{g_\epsilon})- \log \left( 1+\frac{\epsilon}{g_{\min}} \right)\\
    &  \overset{ii}{>}  &
    D_p(\sigma|| \rho_{g_\epsilon}) + 2 \log \left( 1+\frac{\epsilon}{g_{\min}} \right) - \log \left( 1+\frac{\epsilon}{g_{\min}} \right)\\
    &  \overset{iii}{>}  &
    D_p(\sigma|| \rho_g) - \log \left( 1+\frac{\epsilon}{g_{\min}} \right) + 2\log \left( 1+\frac{\epsilon}{g_{\min}} \right) - \log \left( 1+\frac{\epsilon}{g_{\min}} \right)\\
    &  =  &
    D_p(\sigma||\rho_g).
\end{eqnarray*}
\item[]{Case 2: For the range $p \in [1/2,1)$}, we have $\frac{\gamma\;p}{1-p}=\frac{p}{1-p}$. Thus:
\begin{eqnarray*}
    D_p(\rho||\rho_g)
    &  \overset{i}{>}  &
    D_p(\rho||\rho_{g_\epsilon})- \frac{p}{1-p} \log \left( 1+\frac{\epsilon}{g_{\min}} \right)\\
    &  \overset{ii}{>}  &
    D_p(\sigma|| \rho_{g_\epsilon}) + \frac{2p}{1-p}\log \left( 1+\frac{\epsilon}{g_{\min}} \right) - \frac{p}{1-p}\log \left( 1+\frac{\epsilon}{g_{\min}} \right)\\
    &  \overset{iii}{>}  &
    D_p(\sigma|| \rho_g) - \frac{p}{1-p}\log \left( 1+\frac{\epsilon}{g_{\min}} \right) + \frac{2p}{1-p}\log \left( 1+\frac{\epsilon}{g_{\min}} \right) - \frac{p}{1-p}\log \left( 1+\frac{\epsilon}{g_{\min}} \right)\\
    &  =  &
    D_p(\sigma||\rho_g).
\end{eqnarray*}
where for both the above cases, (i) and (iii) follows from Proposition~\ref{prop:continuity_D_alpha}; (ii) follows from hypothesis of the lemma.
\end{itemize}
From analysis of both the cases above, we get that $D_p(\rho||\rho_{g}) > D_p(\sigma||\rho_g)$ which completes the proof of the lemma.
\end{proof}

\noindent For our purpose, it is thus sufficient to choose  $\frac{1}{A_2(\epsilon)}=2^{-\frac{2 \epsilon}{N g_{\min}}}$ which guarantee trumping. 

%%%%%%%%%%%%%%%%%%%%%%%%%%%%%%%%%%%%%%%%%%%%%%%%%%

\subsection{Characterization in terms of $p$ R\'enyi divergence, for $p \in (-\bar s,0)$:} \label{sec:neg_p_thermo}
\noindent In the regime $p \in (-\bar s,0)$, we need to show 
\begin{equation}
    \frac{F_{k,1}\left(\frac{1}{\mathbf x^{\bar s}}\right)}{A_s(\epsilon)} > F_{k,1}\left(\frac{1}{\mathbf y^{\bar s}}\right), \qquad 1\leq k\leq N \ , \nonumber
\end{equation}
implies trumping  for some real valued positive function $A_s(\epsilon)$. To demonstrate this, we need the following lemmas.
%%%%%%%%%%%%%%
\begin{lemma}\label{C3_lemma1}
If $\frac{F_{k,1}\left(\frac{1}{\mathbf x^{\bar s}}\right)}{A_s(\epsilon)} > F_{k,1}\left(\frac{1}{\mathbf y^{\bar s}}\right),$ for some real-valued positive function $A_s(\epsilon)$ with $1\leq k\leq N$ and $\epsilon>0$, then
$$
f_1\left(\frac{1}{\mathbf y^{\bar s}},t\right) < \frac{f_1\left(\frac{1}{\mathbf x^{\bar s}},t\right)}{A_s(\epsilon)} .
$$
\end{lemma}
\begin{proof} 
By Definition~\ref{def:polynomial} of $f_1(x,t)$ and $F_{k,1}(x)$ \cite{klemes2010symmetric}:
    \begin{align*}
        f_1\left(\frac{1}{\mathbf y^{\bar s}},t\right) &= \mathop{\Sigma}\limits_{k}F_{k,1}\left(\frac{1}{\mathbf y^{\bar s}}\right)t^k \\
        &\overset{a}{<} \frac{1}{A_s(\epsilon)}\mathop{\Sigma}\limits_{k}F_{k,1}\left(\frac{1}{\mathbf x^{\bar s}}\right)t^k\\
        & =\frac{f_1\left(\frac{1}{\mathbf x^{\bar s}},t\right)}{A_s(\epsilon)}.
    \end{align*}
   where (a) follows from the hypothesis of the lemma.
\end{proof}

\begin{lemma}\label{C3_lemma2} If $f_1\left(\frac{1}{\mathbf y^{\bar s}},t\right) < \frac{f_1\left(\frac{1}{\mathbf x^{\bar s}},t\right)}{A_s(\epsilon)}$, then for any $\epsilon>0$, $\xi \in (0,1)$ and the choice of $A_s(\epsilon)$ as
\[
    A_s(\epsilon) \geq 2^{\left\{ \frac{\left(1+\frac{\epsilon}{g_{\min}} \right)^{2 (1+ \bar s)}-1}{N}\right\}}
\]
the following holds:
$$\left\|\frac{1}{\mathbf x^{\bar s}}\right\|_{\xi} > \left\|\frac{1}{\mathbf y^{\bar s}}\right\|_{\xi} \left( 1 + \frac{\epsilon}{g_{\min}} \right)^{2 \frac{1+\xi \bar s}{\xi}}\;.$$
\end{lemma}
\begin{proof}
    From Fact~\ref{fact:norm_connects_F}, we have the following:
    \begin{eqnarray} \label{eq:norm_lb_neg}
        N \left\|\frac{1}{\mathbf x^{\bar s}}\right\|_\xi^\xi
        &  =  &
        \frac{1}{I_r(\xi)} \int_0^\infty t^{-\xi-1} \log f_r\left(\frac{1}{\mathbf x^{\bar s}},t\right) dt \notag\\
        &  \overset{a}{>} &
        \frac{1}{I_r(\xi)} \int_0^\infty t^{-\xi-1} \log f_r\left(\frac{1}{\mathbf y^{\bar s}},t\right) dt +
        \log A_s(\epsilon) \int_0^\infty \frac{t^{-\xi-1}}{I_r(\xi)} dt \notag\\
        & \overset{b}{>} &
        N \left\|\frac{1}{\mathbf y^{\bar s}}\right\|_\xi^\xi + \log A_s(\epsilon) \int_{k}^\infty \frac{t^{-\xi-1}}{I_r(\xi)} dt \notag\\
        & = &
        N \left\|\frac{1}{\mathbf y^{\bar s}}\right\|_\xi^\xi \left\{ 1+ N\log A_s(\epsilon) \right\}. 
    \end{eqnarray}
    Here (a) follows from the hypothesis of the lemma whereas in (b) we restrict the range of the definite integral to $\left[k={\left(\xi N^2 \left\|\frac{1}{\mathbf y^{\bar s}}\right\|_\xi^\xi I_r(\xi)\right)}^{-1/\xi},\infty \right]$ instead of considering the full domain $[0,\infty]$, ensuring that it evaluates to $N^2 \left\|\frac{1}{\mathbf y^{\bar s}}\right\|_\xi^\xi$. 
    Now, the choice of $A_s(\epsilon) \geq 2^{\left\{\frac{\left( 1+\frac{\epsilon}{g_{\min}}\right)^{2 (1+\bar s)} -1}{N}\right\} }$ implies:
    \begin{align} \label{eq:simple_neg}
        1+ N\log A_s(\epsilon) &\geq \left( 1+\frac{\epsilon}{g_{\min}}\right)^{2(1+ \bar s)} \overset{i}{>} \left( 1+\frac{\epsilon}{g_{\min}}\right)^{2(1+ \xi\bar s)}
    \end{align}

where (i) holds since $\xi\in (0,1)$. Substituting Eq.~\ref{eq:simple_neg} in Eq.~\ref{eq:norm_lb_neg} and taking the $\xi^{th}$ root both the sides, completes the proof.
\end{proof}

\noindent We now state a lemma that serves as a bridge between $\lVert x \rVert_{-p}$ and $\left\|\frac{1}{\mathbf x}\right\|_{k \bar s}$.
\begin{lemma}\label{C3_lemma3}
$\lVert x \rVert_p < \lVert y \rVert_p \left( 1 + \frac{\epsilon}{g_{\min}} \right)^{2\frac{1-p}{p}}$ ~~ for ~~ $p \in (-\bar s,0)$ $\iff \left\|\frac{1}{\mathbf x^{\bar s}}\right\|_{\xi} > \left\|\frac{1}{\mathbf y^{\bar s}}\right\|_{\xi} \left( 1 + \frac{\epsilon}{g_{\min}} \right)^{2 \frac{1+\xi\bar s}{\xi}} $ ~~for ~~ $\xi \in (0,1)$.
\end{lemma}
\begin{proof}
The equivalence in the lemma can be shown by the following analysis for any $\xi\in(0,1)$:
\begin{align}
\left\|\frac{1}{\mathbf x^{\bar s}}\right\|_{\xi} > \left\|\frac{1}{\mathbf y^{\bar s}}\right\|_{\xi} \left( 1 + \frac{\epsilon}{g_{\min}} \right)^{2 \frac{1+\xi \bar s}{\xi}}  
& \iff \left\|\frac{1}{\mathbf x}\right\|_{\xi\bar s}  > \left\|\frac{1}{\mathbf y}\right\|_{\xi\bar s} \left( 1 + \frac{\epsilon}{g_{\min}} \right)^{2\frac{1+\xi \bar s}{\xi \bar s}}\\
 &\iff \frac{1}{\lVert\mathbf x\rVert_{-\xi \bar s}}> \frac{1}{\lVert \mathbf y\rVert_{-\xi \bar s}} \left( 1 + \frac{\epsilon}{g_{\min}} \right)^{2\frac{1+\xi \bar s}{\xi \bar s}}   \nonumber\\
 &\iff \left\|\mathbf x\right\|_{p} < \left\|\mathbf y\right\|_{p} \left( 1 + \frac{\epsilon}{g_{\min}} \right)^{-2 \frac{1+|p|}{|p|}}, \text{ where } p:=-\xi \bar s \text{ and hence}p\in(-\bar s,0)\\
 &\iff \left\|\mathbf x\right\|_{p} < \left\|\mathbf y\right\|_{p} \left( 1 + \frac{\epsilon}{g_{\min}} \right)^{2 \frac{1-p}{p}}, \quad\text{ for all }p\in(-\bar s,0)\;.\nonumber
 \end{align}
 \end{proof}
\begin{lemma}\label{C3_lemma4}
If $\lVert x \rVert_p < \lVert y \rVert_p \left( 1 + \frac{\epsilon}{g_{\min}} \right)^{2 \frac{1-p}{p}}$ for any given $\epsilon>0$, then:
    $$
    H_p(x)< H_p(y) - 2 \log \left( 1+ \frac{\epsilon}{g_{\min}}\right) .
    $$
\end{lemma}
\begin{proof}
From the definition of R\'enyi $p$-entropy (Eq. 4) and the notation $p=-|p|$ (since $p$ is negative) we have:
    \begin{eqnarray*}
        H_p(x)
        &  =  &
        -\frac{1}{1-p}\log \left( N \times \frac{\mathop{\Sigma}\limits_{i=1}^N x_i^p}{N}\right)\\
        &  =  &
        -\frac{\log N}{1+|p|}+ \frac{|p|}{1+|p|}\log \| x \|_p\\
        &  \overset{a}{<}  &
        -\frac{\log N}{1+|p|} + \frac{|p|}{1+|p|} \log \lVert y \rVert_p + \frac{|p|}{1+|p|} \times \frac{2(1-p)}{p}\log \left( 1+\frac{\epsilon}{g_{\min}}\right)\\
        &  =  &
        -\frac{\log N}{1+|p|} + \frac{|p|}{1+|p|} \log \lVert y \rVert_p - \frac{|p|}{1+|p|} \times \frac{2(1+|p|)}{|p|}\log \left( 1+\frac{\epsilon}{g_{\min}}\right)\\
        &  =  &
        H_p(y) - 2 \log \left( 1+\frac{\epsilon}{g_{\min}}\right),
    \end{eqnarray*}
    where (a) follows from the hypothesis of the lemma.
\end{proof}
\begin{lemma}\label{C3_lemma5}
For any given $\epsilon>0$, if $H_p(x) < H_p(y) -2 \log \left( 1+ \frac{\epsilon}{g_{\min}}\right) $ then:
\begin{eqnarray} 
D_p(\rho||\rho_{g_\epsilon}) & \geq & D_p(\sigma||\rho_{g_\epsilon})+2\log \left( 1+ \frac{\epsilon}{g_{\min}}\right) .
\end{eqnarray}
\end{lemma}
\begin{proof}
The proof essentially follows from the relation between the relative entropy of a given distribution with uniform distribution and the Shannon entropy of a given state, that is, 
     \begin{eqnarray*}
        D_p(\rho||\rho_{g_\epsilon}) 
        & = & 
        D_p(x||u_N) =-\log N - H_p(x)\\
        &  \overset{a}{>}  &
        -\log N - H_p(y) + 2\log \left( 1+ \frac{\epsilon}{g_{\min}}\right)\\
        &  =  &
        D_p(y||u_N) + 2\log \left( 1+ \frac{\epsilon}{g_{\min}}\right)\\
        & = &
        D_p(\sigma||\rho_{g_\epsilon}) +2{p}\log \left( 1+ \frac{\epsilon}{g_{\min}}\right)\;.
    \end{eqnarray*}
Here (a) follows from the hypothesis of the lemma.
\end{proof}
\begin{lemma}\label{C3_lemma6}
    For any given $\epsilon>0$, if $D_p(\rho||\rho_{g_\epsilon}) \geq D_p(\sigma||\rho_{g_\epsilon})+ 2\log \left( 1+ \frac{\epsilon}{g_{\min}}\right)$ then:
$$
D_p(\rho||\rho_g) > D_p(\sigma||\rho_g).
$$
\end{lemma}
\begin{proof}
The proof follows from the following analysis:
    \begin{eqnarray*}
    D_p(\rho||\rho_g)
    & \overset{a}{>} &
    D_p(\rho||\rho_{g_\epsilon}) - \log \left( 1+ \frac{\epsilon}{g_{\min}} \right)\\
    & \overset{b}{>} &
    D_p(\sigma||\rho_{g_\epsilon})+2\log \left( 1+ \frac{\epsilon}{g_{\min}}\right) - \log \left( 1+ \frac{\epsilon}{g_{\min}} \right)\\
    & \overset{c}{>} &
    D_p(\sigma||\rho_{g}) - \log \left( 1+ \frac{\epsilon}{g_{\min}} \right) + 2\log \left( 1+ \frac{\epsilon}{g_{\min}}\right) - \log \left( 1+ \frac{\epsilon}{g_{\min}} \right)\\
    & = &
    D_p(\sigma||\rho_{g}).
    \end{eqnarray*}
    where (a) and (c) follow from Eq.~\ref{eq:cont_p1} of Proposition \ref{prop:continuity_D_alpha}, (b) follows from the hypothesis of the Lemma and (c) follows from Proposition \ref{prop:continuity_D_alpha}.
    Above expression implies $D_p(\rho||\rho_g) > D_p(\sigma||\rho_g)$.
\end{proof}
\subsection{Characterization in terms of $p$ R\'enyi divergence, for $p\in(\overline{r},\infty)$ and $p \in (-\infty, -\bar s]$:} \label{sec:large_p_thermo}
\noindent Following Lemma explores the ranges $p\in(\overline{r},\infty)$ and $p \in (-\infty, -\bar s]$ to guarantee the conditions for trumping.
\begin{lemma}\label{lem:mitraok_thermo}
For a fixed $\epsilon>0$, let $N>1$, $\mathbf x,\mathbf y\in \mathbb R_+^N$, $x_1>y_1 \left( 1+\frac{\epsilon}{g_{\min}}\right)^{2}>0$ and $y_{\min} > x_{\min}\left( 1+\frac{\epsilon}{g_{\min}}\right)^{2}>0$. Let $r$ and $s$ be chosen as:
\begin{equation}
    r=\frac{\log N}{\log x_1-\log y_1\left(1+\frac{\epsilon}{g_{\min}} \right)^2} \quad \text{  and  } \quad s=\frac{\log N}{\log y_{\min}-\log \left\{x_{\min} \left(1+\frac{\epsilon}{g_{\min}} \right)^2\right\}}. \nonumber
\end{equation}
Then, $D_p(\rho||\rho_g) > D_p(\sigma||\rho_g)$ for $\forall p\geq \bar r$ and $\forall p< -\bar s$. 
\end{lemma}
\begin{proof}
Using the choice of $r$ and following the steps outlined in Lemma~\ref{lem:mitraok} we get:
\[
\|\mathbf x\|_p\geq\|\mathbf y\|_p \left(1+\frac{\epsilon}{g_{\min}} \right)^2 \text{ for } p\geq \bar r\;.
\]
Now using the following relation between the $p$-R\'enyi divergence and the our scaled $\ell_p$ norm we get:
\begin{align}\label{eq:D_p_uni}
    D_p(\mathbf x||u_N)=\log N + \frac{p}{p-1} \log \lVert \mathbf x \rVert_p \quad  \Rightarrow \quad D_p(\mathbf x||u_N) > D_p (\mathbf y||u_N) + \frac{2p}{p-1} \log \left(1+\frac{\epsilon}{g_{\min}} \right).
\end{align}
The following analysis leads us to the proof of the lemma for $p> \overline{r}$:
\begin{align} \label{eq:D_p_irr}
    D_p(\rho||\rho_g)&=D_p(\mathbf x||g) \notag\\
    &\overset{a}{\geq}D_p(\mathbf x||g_\epsilon)-\log \left(1+ \frac{\epsilon}{g_{\min}} \right) \notag\\
    &\overset{b}{>}D_p (\mathbf y||g_\epsilon) + \frac{p}{p-1} \log \left(1+\frac{\epsilon}{g_{\min}} \right) \notag\\
    &\overset{c}{>} D_p(\mathbf y||g)=D_p(\sigma||\rho_g)\;. \notag
\end{align}
where (a) and (c) follows from the continuity of $p$-R\'enyi divergence mentioned in Eq.~\ref{eq:cont_1} of Proposition~\ref{prop:continuity_D_alpha}; (b) follows from Eq.~\ref{eq:D_p_uni}.\\
In order to prove the lemma for $p < -\bar{s} $, we make the following observations (for $p<0$):
\[
    \left\lVert \frac{1}{\mathbf x}\right\rVert_p=\frac{1}{\lVert \mathbf x \rVert_{-p}}; \quad H_p(x)=\frac{|p|}{1+|p|} \log \lVert \mathbf x \rVert_{-|p|} \quad \Rightarrow \quad D_p(\mathbf x||u_N)=-\log N - H_p(x) \;.
\]
From Corollary~\ref{cor:2} and the choice of $s$, we get:
\begin{align*}
    &\left\lVert  \frac{1} {\mathbf x} \right\rVert_{-|p|}  > \left\lVert \frac{1} {\mathbf y} \right\rVert_{-|p|} \left( 1+\frac{\epsilon}{g_{\min}}\right)^2\\
    &\Rightarrow \frac{1}{\lVert \mathbf x \rVert_{-|p|}}  > \frac{1}{\lVert \mathbf y \rVert_{-|p|}} \left( 1+\frac{\epsilon}{g_{\min}}\right)^2\\
    &\Rightarrow \lVert \mathbf x \rVert_{-|p|} < \lVert \mathbf y \rVert_{-|p|} \left( 1+\frac{\epsilon}{g_{\min}}\right)^{-2}\\
    &\Rightarrow H_p(\mathbf x ) < H_p(\mathbf y)-\frac{2|p|}{|p|+1} \left( 1+\frac{\epsilon}{g_{\min}}\right)\\
    &\Rightarrow D_p(\mathbf x||u_N ) > D_p(\mathbf y||u_N)+\frac{2|p|}{|p|+1} \left( 1+\frac{\epsilon}{g_{\min}}\right)
\end{align*}
Now doing similar analysis for the case $p>\overline{r}$, gives:
\[
    D_p(\rho||\rho_g)>D_p(\sigma||\rho_g) \text{ for } p \in (-\infty ,  -\bar s]\;.
\]
This completes the proof the lemma.
\end{proof}
%%%%%%%%%%%%%%%%%%%%

\section{Discussion on trumping conditions for general states having coherence}\label{sec:Coherence}
At first, it may be noted that no necessary and sufficient set of conditions are currently known for catalytic state transformations involving states having coherence under the corresponding free operations, rather only necessary conditions have been established~\cite{Lostaglio2015}. Here, we outline a method to derive a finite set of necessary conditions for catalytic majorization in quantum states having coherence.

Before discussing the catalytic majorization conditions for quantum states having coherence, let us first explore the formal aspects of the concept known as {\it Symmetry} \cite{Marvian2014,Lostaglio2015}. This concept has significant applications in physics, and recent insights indicate that any deviation from {\it Symmetry}, termed as {\it asymmetry} serves as a resource for information processing tasks \cite{Marvian2014}.

{\bf Symmetry group and symmetry operation:} For a symmetry group $G$, the symmetry transformation of a density matrix $\rho$ corresponding to the group element $g$ can be represented as follows:
$$g\in G \implies \rho \rightarrow \mathcal{U}_g(\rho) = U_g (\rho) U_g^{\dagger}$$ 
where $U_g$ is unitary.

An evolution $\mathcal{M}$ is said to be symmetric with respect to $G$ if it commutes with the action of the symmetry group i.e., if $[\mathcal{M}, G] = 0$ for all $g \in G$. This implies that the order in which the symmetry transformation and the dynamics occur does not affect the final state. In other words, for every density matrix $\rho$ and each element $g$ in the group $G$, the Equation $\mathcal{M}[\mathcal{U}_g(\rho)] = \mathcal{U}_g[\mathcal{M}(\rho)]$ holds. Similarly, a state is called symmetric if it is invariant under symmetry transformations; otherwise, it is called asymmetric.

Indeed, a resource theory can be formulated where the set of free operations comprises those demonstrating symmetry with respect to the group $G$. Specifically, this theory pertains to quantum coherence between eigenspaces of the observables generating $G$. If the generator is the Hamiltonian $H_s$, the corresponding channel $\mathcal{M}$ is time-translation symmetric.

\begin{definition}
    For any $p \geq 0$, the free coherence of a state $\rho$ with respect to the Hamiltonian $H$ is 
    $$A_p(\rho) := D_p(\rho||\mathcal{N}_H(\rho))$$
    where $\mathcal{N}_H$ is the operation that removes all coherence between energy eigenspaces and $D_p$ is the R\'eyni divergence as defined earlier.
\end{definition}

Similar to how free energies determine the extent to which a state deviates from being thermal, free coherence serves as a measure of how far a state strays from being incoherent in energy. In order to check state transformation between two states in presence of a catalyst, the free energy relations are no longer sufficient rather it is only necessary to go beyond free energy relations to capture the role of quantum coherence in thermodynamical state transformation.

Let us now discuss the conditions for the transformation of states having coherence. For that we borrow and state the theorems from \cite{Lostaglio2015,Lostaglio_review_19}.
\begin{fact} \mbox{\cite[Theorem~1]{Lostaglio2015}}\label{fact1}
    The set of Thermal operations on a quantum state is a strict subset of the set of symmetric quantum operations with respect to the time-translation.
\end{fact}
\begin{fact}\mbox{\cite[Theorem~2]{Lostaglio2015}}
    For all $p\geq 0$, we necessarily have $\Delta A_p \leq 0$ for any thermal operation.
\end{fact}
It is easy to argue that the above fact is true in general for any thermal operation. The R\'enyi divergence follows the relation $ D_p(\mathcal{M}(\rho)||\mathcal{M}(\sigma)) \leq  D_p(\rho|| \sigma)$ for all $p$. Since from Fact \ref{fact1}, $[\mathcal{M},\mathcal{N}_H] = 0$, we derive the condition that the above fact is indeed true.\\
\\
A state $\rho$ in $\mathcal{H}$ can be transformed into a state $\sigma$ (i.e., $\rho \rightarrow \sigma$) through a catalytic thermal operation, if there is another quantum state $\rho_c$ in Hilbert space $\mathcal{H}_c$ with Hamiltonian $H_c$ and a thermal operation $\mathcal{M}$ on $\mathcal{H}\otimes \mathcal{H}_c$ i.e.,
$$\mathcal{M}(\rho \otimes \rho_c) = \sigma \otimes \rho_c.$$ 
\begin{fact}\mbox{\cite[Theorem~4]{Lostaglio2015}}
Catalytic thermal operations with a block-diagonal catalyst are symmetric operations, i.e., if $H$ is the system's Hamiltonian and $\mathcal{C}$ is a catalytic thermal operation then
$$\mathcal{C}(e^{-iHt}\rho e^{iHt}) = e^{-iHt} \mathcal{C}(\rho) e^{iHt}.$$
\end{fact}

\begin{fact}\mbox{\cite[Theorem~5]{Lostaglio2015}}
If $[\rho_c,H_c]=0$, for $\rho \otimes \rho_c \rightarrow \sigma \otimes \rho_c$, we necessarily have 
$$A_p(\sigma) \leq A_p (\rho) ~~\forall p \geq 0.$$    
\end{fact}
This implies the necessary conditions for catalytically transforming states having coherence, involve comparing the generalized coherence-free energies (of order $p$) of $\rho$ and $\sigma$ across an infinite range of $p$ values.

One can now get a finite number of necessary conditions for catalytic majorization by replacing $\rho_g$ with a state $\rho_f=:\mathcal{N}_H(\rho)$ that removes all the off diagonal elements of the state into the discussion of Corollary~\ref{cor:second_law_rational} and Theorem~\ref{thm:second_law}.

\end{document}